\documentclass[fleqn,usenatbib]{mnras}

\usepackage{newtxtext,newtxmath}

\usepackage[T1]{fontenc}

\DeclareRobustCommand{\VAN}[3]{#2}
\let\VANthebibliography\thebibliography
\def\thebibliography{\DeclareRobustCommand{\VAN}[3]{##3}\VANthebibliography}

\usepackage{graphicx}	
\usepackage{amsmath}	
\usepackage{CJKutf8}
\usepackage{dcolumn}
\usepackage{gensymb}
\usepackage{graphicx}	
\usepackage{hyperref}   
\usepackage{multirow}

\newcolumntype{d}{D{.}{.}{6}}
\newcommand{\anote}{^{\mathrm{a}}}              
\newcommand{\avgux}{\langle u_x\rangle}         
\newcommand{\avguz}{\langle u_z\rangle}         
\newcommand{\avgvx}{\langle v_x\rangle}         
\newcommand{\avgvz}{\langle v_z\rangle}         
\newcommand{\bnote}{^{\mathrm{b}}}              
\newcommand{\cnote}{^{\mathrm{c}}}              
\newcommand{\cs}{c_\mathrm{s}}                  
\newcommand{\dnote}{^{\mathrm{d}}}              
\newcommand{\dPdrhog}{\mathrm{dP} / \od\rhog}   
\newcommand{\dPduxz}{\mathrm{dP} / \od u_{x,z}} 
\newcommand{\dPdux}{\mathrm{dP} / \od u_x}      
\newcommand{\dPduz}{\mathrm{dP} / \od u_z}      
\newcommand{\dPdvxz}{\mathrm{dP} / \od v_{x,z}} 
\newcommand{\dPdvx}{\mathrm{dP} / \od v_x}      
\newcommand{\dPdvz}{\mathrm{dP} / \od v_z}      
\newcommand{\Dpscale}{D_\mathrm{p,scale}}       
\newcommand{\Dpx}{D_{\mathrm{p}_x}}             
\newcommand{\Dpz}{D_{\mathrm{p}_z}}             
\newcommand{\Dpxz}{D_{\mathrm{p}_{x,z}}}        
\newcommand{\drhog}{\delta\rho_\mathrm{g}}      
\newcommand{\etavK}{\eta v_\mathrm{K}}          
\newcommand{\Hg}{H_\mathrm{g}}                  
\newcommand{\Hp}{H_\mathrm{p}}                  
\newcommand{\Ma}{\mathrm{Ma}}                   
\newcommand{\od}{\mathrm{d}}                    
\newcommand{\OmegaK}{\Omega_\mathrm{K}}         
\newcommand{\Rg}{\mathcal{R}_\mathrm{g}}        
\newcommand{\rhog}{\rho_\mathrm{g}}             
\newcommand{\rhogn}{\rho_{\mathrm{g},0}}        
\newcommand{\rhop}{\rho_\mathrm{p}}             
\newcommand{\rhopCDF}{\mathrm{P}(>\rhop)}       
\newcommand{\Rp}{\mathcal{R}_\mathrm{p}}        
\newcommand{\St}{\mathrm{St}}                   
\newcommand{\taue}{\tau_\mathrm{e}}             
\newcommand{\taus}{\tau_\mathrm{s}}             
\newcommand{\tlim}{t_\mathrm{lim}}              
\newcommand{\teddy}{t_\mathrm{eddy}}            
\newcommand{\tstop}{t_\mathrm{stop}}            
\newcommand{\uvec}{\mathbf{u}}                  
\newcommand{\vvec}{\mathbf{v}}                  
\newcommand{\Zeff}{Z_\mathrm{eff}}              

\title[Streaming instability and pressure gradient]{Dust--gas dynamics driven by the streaming instability with various pressure gradients}

\author[S. A. Baronett et al.]{
Stanley A. Baronett,$^{1,2}$\thanks{E-mail: \href{mailto:barons2@unlv.nevada.edu}{barons2@unlv.nevada.edu} (SAB)}
\begin{CJK*}{UTF8}{bkai}
Chao-Chin Yang (楊朝欽),$^{1,2,3}$
\end{CJK*}
\begin{CJK*}{UTF8}{bkai}
Zhaohuan Zhu (朱照寰)$^{1,2}$
\end{CJK*}
\\
$^{1}$Department of Physics and Astronomy,
University of Nevada, Las Vegas,
Box 454002,
4505 S. Maryland Pkwy.,
Las Vegas, NV 89154-4002, USA\\
$^{2}$Nevada Center for Astrophysics,
University of Nevada, Las Vegas,
4505 S. Maryland Pkwy.,
Las Vegas, NV 89154-4002, USA\\
$^{3}$Department of Physics and Astronomy,
The University of Alabama,
Box 870324,
Tuscaloosa, AL 35487-0324, USA
}

\date{Accepted 2024 January 18. Received 2024 January 17; in original form 2023 August 21}

\pubyear{2015}

\begin{document}
\label{firstpage}
\pagerange{\pageref{firstpage}--\pageref{lastpage}}
\maketitle

\begin{abstract}
The streaming instability, a promising mechanism to drive planetesimal formation in dusty protoplanetary discs, relies on aerodynamic drag naturally induced by the background radial pressure gradient.
This gradient should vary in disks, but its effect on the streaming instability has not been sufficiently explored.
For this purpose, we use numerical simulations of an unstratified disc to study the non-linear saturation of the streaming instability with mono-disperse dust particles and survey a wide range of gradients for two distinct combinations of the particle stopping time and the dust-to-gas mass ratio.
As the gradient increases, we find most kinematic and morphological properties increase but not always in linear proportion.
The density distributions of tightly-coupled particles are insensitive to the gradient whereas marginally-coupled particles tend to concentrate by more than an order of magnitude as the gradient decreases.
Moreover, dust--gas vortices for tightly-coupled particles shrink as the gradient decreases, and we note higher resolutions are required to trigger the instability in this case.
In addition, we find various properties at saturation that depend on the gradient may be observable and may help reconstruct models of observed discs dominated by streaming turbulence.
In general, increased dust diffusion from stronger gradients can lower the concentration of dust filaments and can explain the higher solid abundances needed to trigger strong particle clumping and the reduced planetesimal formation efficiency previously found in vertically-stratified simulations.
\end{abstract}




\section{Introduction}
\label{sec:introduction}

In dusty protoplanetary discs, sub-micron interstellar dust grains must grow by 13 orders of magnitude in size to become fully-grown planets, albeit the details of various stages in the planet formation process remain difficult to disentangle.
In these stages, the formation of planetesimals plays a key role, bridging dust coagulation and the formation of planetary cores \citep[and references within]{JohansenBlumTanaka2014, BirnstielFangJohansen2016}.

One mechanism to drive the formation of planetesimals is the streaming instability, first discovered by \cite{YoudinGoodman2005}.
The development of this instability is composed of four stages: linear growth \citep{YoudinJohansen2007, KrappBenitez-LlambayGressel2019, PaardekooperMcNallyLovascio2020, ZhuYang2021}, non-linear saturation \citep{JohansenYoudin2007, YangZhu2021}, vertical sedimentation \citep{YangJohansen2014, LiYoudinSimon2018}, and strong clumping \citep{JohansenYoudinMacLow2009}.
Dependent on the grain size, a vertically-equilibrated dust layer needs to reach a vertically-integrated solid abundance threshold to trigger the strong clumping of dust materials in the mid-plane of the disc \citep{CarreraJohansenDavies2015, YangJohansenCarrera2017, LiYoudin2021}.
This strong clumping thus produces the particle concentrations needed for gravitational collapse into planetesimals \citep{JohansenMacLowLacerda2015, SimonArmitageLi2016, SchaferYangJohansen2017, NesvornyLiSimon2021}.
Since this instability relies on aerodynamic drag, it fundamentally depends on a difference in velocity between the solid particles and the surrounding gas.
An important and natural source leading to this velocity difference is the background radial pressure gradient of the gaseous disc \citep{AdachiHayashiNakazawa1976, Weidenschilling1977MNRAS}.

In both theoretical models and observations, the pressure gradient of the gas within a protoplanetary disc can vary radially.
Although standard discs derived from a viscous disc evolution model \citep{Lynden-BellPringle1974, Hartmann1998} or the minimum-mass Solar Nebula \citep{Weidenschilling1977ApSS, Hayashi1981} traditionally feature a smooth pressure profile and thus a slowly varying gradient \citep{BaiStone2010}, theoretical models incorporating magnetic fields and more realistic thermodynamics (e.g. ice lines) commonly develop large-scale, long-lived axisymmetric pressure variations and extrema \citep{LyraJohansenKlahr2008, JohansenYoudinKlahr2009, KretkeLinGaraud2009, LyraJohansenZsom2009, DzyurkevichFlockTurner2010, SimonBeckwithArmitage2012, DittrichKlahrJohansen2013, BaiStone2014, SimonArmitage2014, BitschJohansenLambrechts2015, BethuneLesurFerreira2017}.
Moreover, anisotropic infall of filamentary accretion streams onto embedded discs from the star-forming environment can generate vortices and azimuthal shear which can also form robust pressure maxima \citep{KuznetsovaBaeHartmann2022}.
Meanwhile, disc observations of the (sub-)millimetre continuum emission at high angular resolutions \citep{ALMAPartnershipBroganPerez2015, AndrewsHuangPerez2018} have identified bright rings, suggesting the dust could be trapped within gas pressure maxima \citep{Whipple1972, Whipple1973}.
Furthermore, recent molecular-line observations of discs \citep[][e.g. HD 163296]{ObergGuzmanWalsh2021} reveal substructures in the gas with large-scale pressure variations.

The effect of the radial pressure gradient on the streaming instability has not been well studied, although a couple of investigations into it have already identified significant consequences.
For a given particle size distribution, \cite{BaiStone2010L} found the critical solid abundance needed to trigger strong particle clumping increases monotonically with the pressure gradient.
For a single particle size with self-gravity, \citet[\S~3.1.1]{AbodSimonLi2019} found the planetesimal formation efficiency steeply decreases as the gradient increases.
Although both groups attribute their results to an increase in turbulence driven by the streaming instability as the gradient increases, neither one provided any quantitative analysis of the turbulent properties of their systems.
Therefore, we aim in this work to quantify the dependence of turbulence on the radial pressure gradient by using vertically-unstratified models to better bridge the gap between mid-plane turbulence and vertical sedimentation, in the development of the streaming instability.

Here, we introduce the first thorough investigation into the non-linear saturation of the streaming instability with various pressure gradients in numerical simulations of an unstratified disc.
In Section~\ref{sec:methodology}, we describe the equations governing the gas and the dust, our numerical methods, and our model setup.
In Section~\ref{sec:results}, we detail the saturation state, morphology, and kinematics resulting from our models.
In Section~\ref{sec:discussion}, we discuss the implications of our findings for planetesimal formation and radial transport and compare our results with recent observations.
We conclude with a summary in Section~\ref{sec:conclusions}.

\section{Methodology}
\label{sec:methodology}

We model a system of gas and solid particles using the local-shearing-box approximation \citep{GoldreichLynden-Bell1965}.
The computational domain is at an arbitrary distance $r$ from the central star and revolves around the star at the local Keplerian angular frequency $\OmegaK$.
The domain is small compared with the orbital distance, and the equations of motion can be linearised such that the domain approximates a rectangular box with its $x$, $y$, and $z$ axes constantly aligned in the radial, azimuthal, and vertical directions, respectively.
As we assume the system is axisymmetric and omit the vertical component of stellar gravity in this work, the boundary conditions are thus periodic in both $x$ and $z$ directions.
For simplicity, we omit the effects of magnetic fields.
The following Subsections~\ref{sec:gas}, \ref{sec:dust}, \ref{sec:numerical_method}, and \ref{sec:model_setup} detail the equations of motion for the fluid gas, those for the solid particles, the numerical method we use to solve them, and our model setup, respectively.

\subsection{Gas}
\label{sec:gas}

The continuity and momentum equations for the gas are
\begin{equation}
    \frac{\partial\rhog}{\partial t} + \nabla\cdot(\rhog\uvec) =
    0,
    \label{eq:gas_cont}
\end{equation}
\begin{align}
    \label{eq:gas_mom}
    &\frac{\partial\rhog\uvec}{\partial t} + \nabla\cdot(\rhog\uvec\uvec + P\mathbf{I})\notag\\&= \rhog\left[2\OmegaK u_y\hat{\mathbf{x}} - \frac{1}{2}\OmegaK u_x\hat{\mathbf{y}} + 2\OmegaK\Pi \cs\hat{\mathbf{x}} - \frac{\rhop}{\rhog}\left(\frac{\uvec - \mathbf{v}}{\tstop}\right)\right],
\end{align}
respectively.
We solve for the gas density $\rhog$ and the gas velocity $\uvec$ with the velocity measured relative to the background Keplerian shear flow $\uvec^\prime = -(3/2)\OmegaK x\hat{\mathbf{y}}$.
The pressure $P$ is given by the isothermal equation of state $P=\rhog\cs^2$, where $\cs$ is the speed of sound, and $\mathbf{I}$ is the identity matrix.
The first two source terms on the right-hand side of equation~\eqref{eq:gas_mom} are a combination of the radial component of the stellar gravity and the Coriolis and the centrifugal forces.
The third term is a constant outward force on the gas due to an external radial pressure gradient, determined by the dimensionless parameter \citep[][eq.~1]{BaiStone2010}
\begin{equation}
    \Pi\equiv\frac{\etavK}{\cs}=\frac{\eta r}{\Hg},
	\label{eq:Pi}
\end{equation}
where $v_\mathrm{K}$ is the local Keplerian velocity, $\Hg = \cs / \OmegaK$ is the vertical gas scale height, and
\begin{equation}
    \eta \equiv -\frac{1}{2}\frac{1}{\rhog\OmegaK^2r}\frac{\partial P}{\partial r} = -\frac{1}{2}\left(\frac{\Hg}{r}\right)^2\frac{\partial\ln P}{\partial\ln r} \sim \left(\frac{\cs}{v_\mathrm{K}}\right)^2,
	\label{eq:eta}
\end{equation}
is the fractional reduction in orbital speed of the gas from Keplerian (when $\eta > 0$) if the dust were not present \citep[][eq.~1.9]{NakagawaSekiyaHayashi1986}.
The fourth and final term is the frictional drag force from the solid particles back to the gas, where $\mathbf{v}$ is the ensemble-averaged local velocity of the particles, again measured relative to the background shear, and $\tstop$ is the stopping time (Section~\ref{sec:dust}).
The factor of the dust-to-gas density ratio $\rhop / \rhog$ ensures the conservation of the total linear momentum of the gas and dust particles, where $\rhop$ is the averaged dust density in the gas cell.

We initialise the gas as follows.
The gas density field is initially uniform with $\rhog(x,y,z,t=0) = \rho_\mathrm{g,0}$.
By assuming a total dust-to-gas mass ratio
\begin{equation}
    \epsilon \equiv \frac{\langle\rhop\rangle}{\rhogn},
	\label{eq:epsilon}
\end{equation} where
\begin{equation}
    \langle f \rangle \equiv \frac{1}{L_x L_y L_z} \iiint f \od x\od y\od z
	\label{eq:vol-avg}
\end{equation}
is the instantaneous volume average of quantity $f$ over the computational domain of dimensions $L_x\times L_y\times L_z$, we then uniformly apply the equilibrium solution by \citet{NakagawaSekiyaHayashi1986} to the radial and azimuthal components of the gas velocity, while setting the vertical component to zero.

Our primary objective is to study the effects of the radial pressure gradient on dust--gas dynamics driven by the streaming instability.
While $\Pi = 0.05$ is considered typical in the inner regions of a wide range of disc models \citep{BaiStone2010, BitschJohansenLambrechts2015}, the radial pressure gradient can vary locally due to disc substructures or globally depending on location and the evolutionary stage of the disc (Section~\ref{sec:introduction}).
Thus, we study four values of $\Pi$ -- 0.01, 0.02, 0.05, and 0.1 -- that cover a typical range of the gradient (Table~\ref{tab:params}).

\subsection{Dust}
\label{sec:dust}

Following \cite{YoudinJohansen2007}, we model the dust as Lagrangian super-particles, each of which represents an ensemble of numerous identical solid particles described by their total mass and average velocity.
The equations of motion for the $i$-th super-particle is then
\begin{align}
    \label{eq:dust_vel}
    \frac{\od\mathbf{x}_{\mathrm{p},i}}{\od t} &= \mathbf{v}_i - \frac{3}{2}\OmegaK x_{\mathrm{p},i} \hat{\mathbf{y}},\\
    \label{eq:dust_acc}
    \frac{\od\mathbf{v}_i}{\od t} &= 2\OmegaK v_{i,y}\hat{\mathbf{x}} - \frac{1}{2}\OmegaK v_{i,x}\hat{\mathbf{y}} - \frac{\mathbf{v}_i - \uvec}{\tstop},
\end{align}
where the velocity $\mathbf{v}_i$ is measured relative to the background Keplerian shear $\mathbf{v}_i^\prime = -(3/2)\OmegaK x_{\mathrm{p},i}\hat{\mathbf{y}}$.
The gas velocity $\uvec$ is evaluated at the particle position $\mathbf{x}_{\mathrm{p},i}$ by interpolation (Section~\ref{sec:numerical_method}).
The right-hand side of equation~\eqref{eq:dust_acc} parallels equation~\eqref{eq:gas_mom} in Lagrangian form, except for the absence of the radial gas pressure gradient.

The stopping time $\tstop$ in equations~\eqref{eq:gas_mom} and \eqref{eq:dust_acc} is the $e$-folding time to damp the relative speed between a solid particle and the surrounding gas due to their mutual frictional drag \citep{Whipple1972, Weidenschilling1977MNRAS}.
We assume all particles have the same stopping time (mono-disperse dust particles), and the self-gravity and collisions between them are ignored.
As was done by \cite{YoudinGoodman2005}, we use the dimensionless stopping time (a.k.a.\ Stokes number)
\begin{equation}
    \taus \equiv \OmegaK \tstop.
    \label{eq:taus}
\end{equation}
This parameter measures how well a particle is coupled to the gas: the smaller the $\taus$, the tighter the coupling.

We initialise the dust particles as follows.
We use a total number of particles such that there are $n_\mathrm{p}=4$ particles per cell on average, and we randomly distribute these particles throughout the domain.
Since the particles are identical, the mass of each particle can be found given the total solid-to-gas density ratio $\epsilon$ defined by equation~\eqref{eq:epsilon}.
Similar to the gas, we uniformly apply the equilibrium solution by \citet{NakagawaSekiyaHayashi1986} to the radial and azimuthal components of the particle velocity, while setting the vertical component to zero.

\subsection{Numerical method}
\label{sec:numerical_method}

To simultaneously solve equations~\eqref{eq:gas_cont}, \eqref{eq:gas_mom}, \eqref{eq:dust_vel}, and \eqref{eq:dust_acc}, we use \textsc{athena++} \citep{StoneTomidaWhite2020}, a modular and parallelized astrophysical magnetohydrodynamics code.
The gas is solved by the finite volume method, and we use the HLLE Riemann solver, a Courant number of 0.4, the second-order van Leer predictor--corrector time-integration scheme, and the piecewise parabolic method applied to primitive variables for spatial reconstruction.

We have extended the code to simultaneously simulate dust grains as Lagrangian super-particles (Yang et al. in preparation), each with an individual position and velocity that are integrated in unison with the hydrodynamic time-steps.
To model their interaction with the Eulerian gas, we employ the standard particle--mesh method \citep{HockneyEastwood1981} using the Triangular-Shaped-Cloud scheme to interpolate the gas properties to the particles and assign the particle properties to the mesh with high spatial accuracy.

\subsection{Model setup}
\label{sec:model_setup}

We limit our survey to two distinct cases, each with a different combination of the dimensionless stopping time $\taus$, defined by equation~\eqref{eq:taus}, and the dust-to-gas mass ratio $\epsilon$, defined by equation~\eqref{eq:epsilon}, that were first studied by \cite{JohansenYoudin2007} and represent contrasting non-linear regimes of the streaming instability.
Case~AB has $\taus = 0.1$ and $\epsilon = 1.0$, while Case~BA has $\taus = 1.0$ and $\epsilon = 0.2$.
In other words, Case~AB contains an enhanced abundance of relatively tightly-coupled particles, while Case~BA contains a relatively low abundance of marginally-coupled particles.
For compact solid particles at $r\approx10$~au (e.g. at the semimajor axis of Saturn) with a uniform density of 1 g cm$^{-3}$, $\taus = 0.1$ and 1.0 roughly correspond to particles of 2~cm and 20~cm in size, respectively, in standard minimum mass solar nebular models \citep[][fig.~3]{JohansenBlumTanaka2014}.
The domain sizes are $L_x\times L_z = 0.1\,\Hg\times0.1\,\Hg$ and $2\,\Hg\times2\,\Hg$ for the AB and BA~cases, respectively.
Based on linear growth rate maps \citep[fig.~1]{YoudinJohansen2007}, for Case~AB, this allows 47 wavelengths of the fastest-growing mode when $\Pi = 0.01$ and 4 when $\Pi = 0.1$, and for Case~BA, this allows 31 when $\Pi = 0.01$, and 3 when $\Pi = 0.1$.
These two cases when $\Pi = 0.05$ can also serve as comparisons with previous works \citep{JohansenYoudin2007, BaiStone2010, Benitez-LlambayKrappPessah2019}.

As in \citet[][\S~3.1]{JohansenYoudin2007}, 
the randomly-distributed particles (Section~\ref{sec:dust}) provides a white-noise power spectrum in the particle density field to seed the streaming instabilities.
We run the simulations of these two cases well into the saturation stage for statistical analysis of the state.
The AB and BA~cases are run for $t_\mathrm{lim} = 10T$ and $200T$, respectively, where $T = 2\pi / \OmegaK$ is the local orbital period.
We report our results at our highest resolution of $2048\times2048$ cells, and we leave the resolution study in Appendix~\ref{sec:resolution_study}.
Table~\ref{tab:params} summarises the parameters for each of our eight runs.

\begin{table}
	\centering
	\caption{Simulation model parameters.
	The columns are (1) case name, (2) dimensionless stopping time$\anote$, (3) total dust-to-gas mass ratio$\bnote$, (4) domain size, (5) simulation time limit, and (6) dimensionless radial pressure gradient$\cnote$.
	Length and time are in units of gas scale height $\Hg$ and orbital period $T$, respectively.}
	\label{tab:params}
	\begin{tabular}{ccccrc} 
        \hline
        Case & $\taus$ & $\epsilon$ & $L_x = L_z$ & \multicolumn{1}{c}{$\tlim$} & \multicolumn{1}{c}{$\Pi$}              \\
             &         &            & ($\Hg$)     & \multicolumn{1}{c}{($T$)}   &                                        \\
        (1)  & (2)     & (3)        & (4)         & (5)                         & (6)                                    \\
        \hline
        AB   & 0.1     & 1.0        & 0.1         & 10                          & \multirow{2}{*}{0.01, 0.02, 0.05, 0.1} \\
        BA   & 1.0     & 0.2        & 2.0         & 200                         &                                        \\
        \hline
        \multicolumn{3}{l}{$\anote$ Defined by equation~\eqref{eq:taus}}\\
        \multicolumn{3}{l}{$\bnote$ Defined by equation~\eqref{eq:epsilon}}\\
        \multicolumn{3}{l}{$\cnote$ Defined by equation~\eqref{eq:Pi}}\\
	\end{tabular}
\end{table}

\section{Results}
\label{sec:results}

We report and analyse the results of our simulations as follows.
Subsection~\ref{sec:saturation_state} covers dispersion properties for each of our runs and how we determine the saturation state.
Subsection~\ref{sec:morphology} details the saturation morphology of the gas and dust density fields.
Finally, we discuss kinematic details of the gas and dust in subsection~\ref{sec:kinematics}.

\subsection{Saturation state}
\label{sec:saturation_state}

\begin{figure*}
	\includegraphics[width=\textwidth]{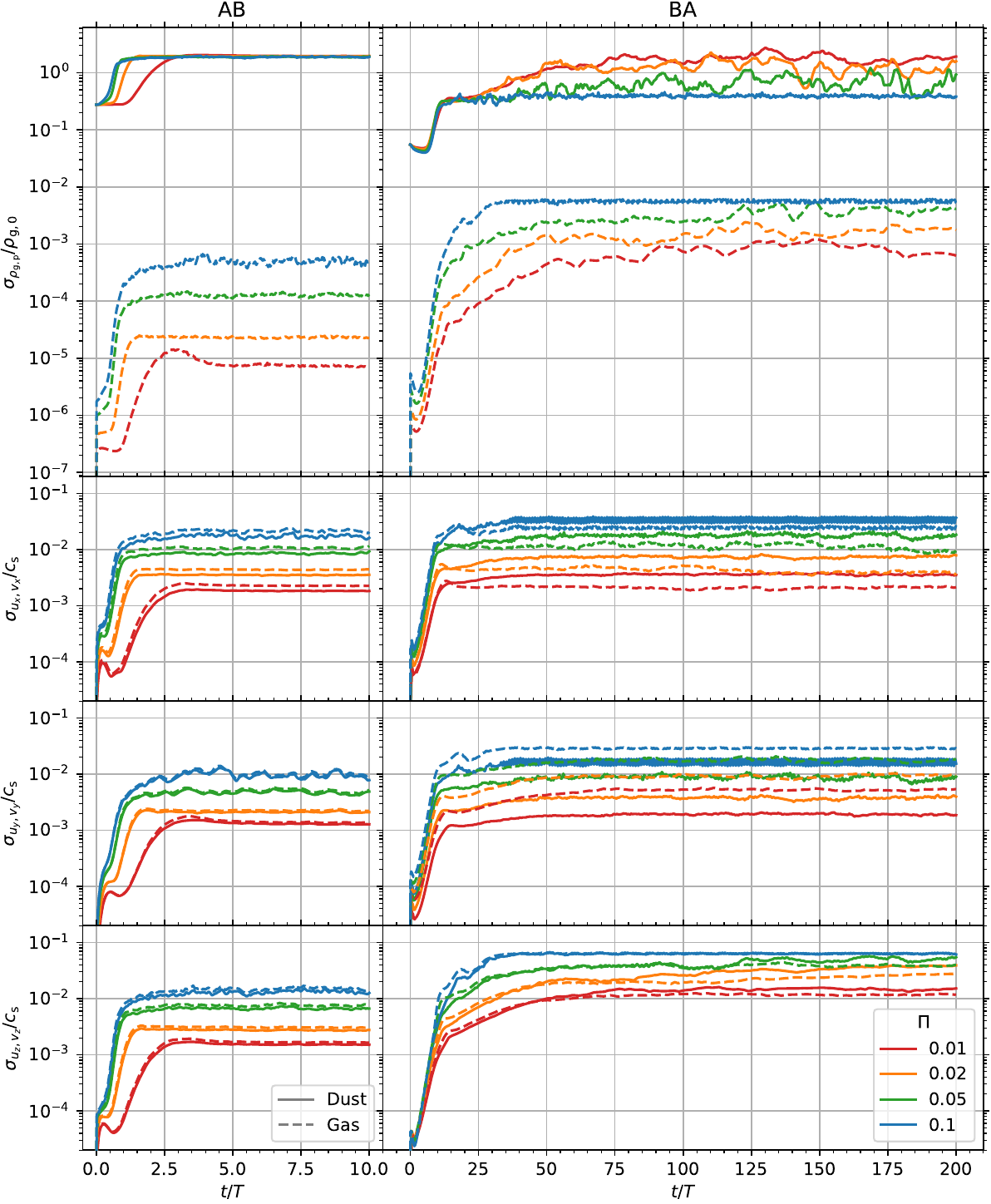}
    \caption{Dispersions as a function of time for all models in Table~\ref{tab:params}.
    The left and right columns are for Cases~AB ($\taus = 0.1$ with $\epsilon = 1.0$) and BA ($\taus = 1.0$ with $\epsilon = 0.2$), respectively.
    The rows from top to bottom show the densities, defined by equation~\eqref{eq:gas_density_dispersion}, and the three components of the velocities, defined by equation~\eqref{eq:gas_velocity_dispersion}.
    Solid and dashed lines represent the dust and the gas, respectively, while line colours represent models with different values of the dimensionless radial pressure gradient $\Pi$.
    Densities and velocities are normalised to the initially uniform gas density $\rhogn$ and the speed of sound $\cs$, respectively.}
    \label{fig:dispersions}
\end{figure*}

Before we can analyse the dust--gas dynamics driven by the streaming instability, we must first determine when each of our models transitions from linear growth to non-linear saturation.
Thus, we construct and plot the time evolution of several statistical diagnostics.
We define the time-dependent gas density dispersion by
\begin{equation}
    \sigma_{\rhog} \equiv \sqrt{\langle\drhog^2\rangle - \langle\drhog\rangle^2},
	\label{eq:gas_density_dispersion}
\end{equation}
where the notation $\langle \cdot \rangle$ is the volume average defined by equation~\eqref{eq:vol-avg}, and $\drhog \equiv \rhog - \langle\rhog\rangle$ is the local gas density deviation from the mean gas density $\langle\rhog\rangle = \rhogn$.
Similarly, we define the gas velocity dispersion for each component as
\begin{equation}
    \sigma_{u_{x,y,z}} \equiv \sqrt{\frac{\langle\rhog\delta u_{x,y,z}^2\rangle}{\rhogn} - \Delta u_{x,y,z}^2},
    \label{eq:gas_velocity_dispersion}
\end{equation}
where
\begin{equation}
    \Delta\uvec \equiv \frac{\langle\rhog\delta\uvec\rangle}{\rhogn}
    \label{eq:gas-velocity-deviation}
\end{equation}
is the mass-weighted average gas velocity deviation from the initial equilibrium velocity $\uvec_0$, and $\delta\uvec \equiv \uvec - \uvec_0$.
As the dust is comprised of Lagrangian super-particles, we first map particle properties $\rhop$ and $\vvec$ to the gas grid via the particle-mesh assignment.
Then we compute the dispersions of the density $\sigma_{\rhop}$ and velocity components, e.g. $\sigma_{v_x}$, of the dust via equations in parallel with \eqref{eq:gas_density_dispersion} and \eqref{eq:gas_velocity_dispersion}, respectively.
Fig.~\ref{fig:dispersions} shows the evolution of density and velocity dispersions of the gas and dust for each case and value of $\Pi$ in Table~\ref{tab:params}.

For Case~AB, all diagnostics reach a quasi-steady level, indicating saturation.
Compared to models with lower values of $\Pi$, those with higher values show faster rates of increase prior to this state and reach saturation earlier.
\citet[][\S~3.3.1 and fig.~5]{JohansenYoudin2007} described the early growth stage of their Run~AB for $\Pi = 0.05$, which can also be seen in our available videos (see `Data Availability' section).
In particular, they suggested the rapid appearance and non-uniform growth of cavities (i.e., voids with dense inner rims) within $t \approx 1T$ can be explained by local Poisson fluctuations around $\rhop/\rhog = 1$ (value of $\epsilon$, as defined by equation~\eqref{eq:epsilon}, in Case~AB), where the peak linear growth rate of the streaming instability for $\taus = 0.1$ sharply rises by an order of magnitude \citep[fig.~1]{JohansenYoudin2007}, noting that the corresponding growth time-scale for $\rhop/\rhog = 1.25$ is half that for $\rhop/\rhog = 1.0$.\footnote{\citet[fig.~2]{YoudinJohansen2007} showed the associated phase speed of the fastest growing radial waves for $\taus = 0.1$ changes sign from inward to outward for $\rhop/\rhog \lessapprox 1$ and $\rhop/\rhog \gtrapprox 1$, respectively, coinciding with this steep increase in the peak linear growth rate.}
By carefully seeding a linear mode to suppress all Poisson noise in one variation of their Run~AB, they found numerical growth matched the peak analytic rate for $\rhop/\rhog = 1$ \citep[][fig.~6]{JohansenYoudin2007}.
Moreover, by lowering the amplitude of Poisson fluctuations with twice as many particles per grid cell in another variation, they found a delay in the appearance of cavities.
While their analysis seems to support their explanation for the rapid development of cavities as another manifestation of the streaming instability, we cannot yet rule out the initial presence of, or subsequent turbulence driven by, some secondary instabilities in Case~AB.

\begin{figure*}
	\includegraphics[width=\textwidth]{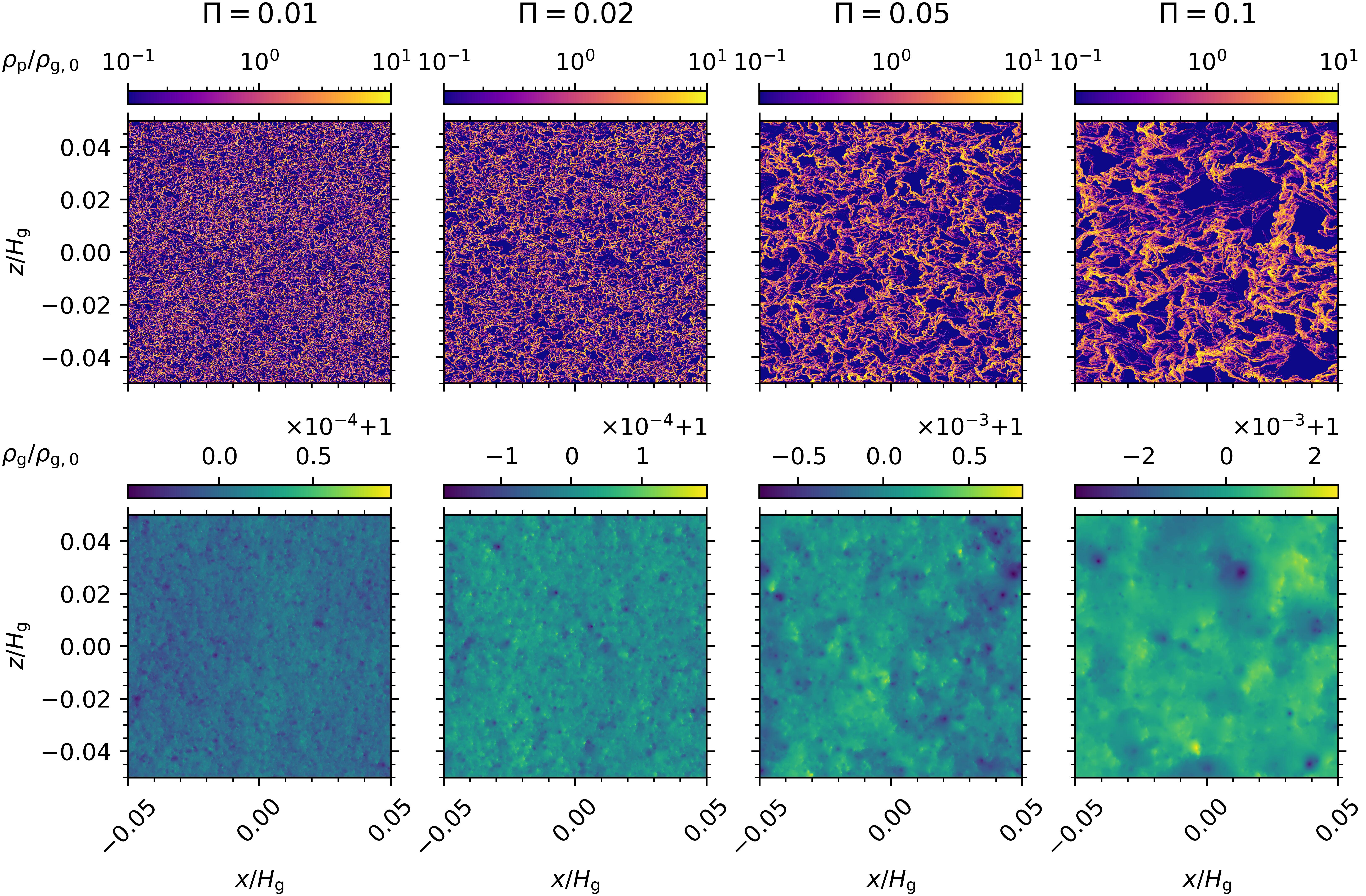}
    \caption{Final snapshots at $\tlim = 10T$ of the density fields during the saturation state of Case~AB ($\taus = 0.1$ with $\epsilon = 1.0$).
    The top and bottom rows show the dust and gas density fields, respectively, while columns from left to right show models with increasing radial pressure gradient $\Pi$.}
    \label{fig:AB_snapshots}
\end{figure*}

Judging from Fig.~\ref{fig:dispersions}, we average all reported quantities for the saturation state of all AB~models from $t = 5T$ to $\tlim = 10T$ throughout this work, unless otherwise specified.
Table~\ref{tab:time_averages} lists the time-averaged dispersions from Fig.~\ref{fig:dispersions}.
We note the dust density dispersion $\sigma_{\rhop}$ for Case~AB is on the order of 2$\rhogn$ and, more importantly, is rather insensitive to $\Pi$, with a relative difference of at most 5\% between models.
By contrast, the gas density dispersion $\sigma_{\rhog}$ significantly increases as $\Pi$ increases, differing by as much as two orders of magnitude between $\Pi = 0.01$ and $0.1$ models.
Despite this large span, the strongest density fluctuation of all our AB~models only reaches the order of $10^{-4}\rhogn$, indicating the gas remains relatively incompressible under the streaming instability \citep{JohansenYoudin2007, YangZhu2021}.
As for the velocity dispersions of the gas and the dust, $\sigma_{\uvec}$ and $\sigma_{\vvec}$, we find these also increase as $\Pi$ increases.
Yet, for any given $\Pi$, $\sigma_{u_x,v_x} > \sigma_{u_z,v_z} > \sigma_{u_y,v_y}$, similar to the relationship found for multi-species Model~Af by \cite{YangZhu2021}.
Moreover, we find $\sigma_{u_x} \approx 2\sigma_{u_y}$, consistent with gas epicyclic motions \citep{PapaloizouTerquem2006, YangMacLowMenou2009}, except perhaps for $\Pi = 0.01$ where $\sigma_{u_x} \approx 1.6\sigma_{u_y}$.
As shown in Fig.~\ref{fig:dispersions} and Table~\ref{tab:time_averages}, the time variability also increases as $\Pi$ increases, except for $\sigma_{\rhop}$ between $\Pi = 0.01$ and $0.02$, but only amounts to at most 10\% for $\Pi = 0.1$.
The final dust density fields at $\tlim = 10T$ for each AB~model is shown in Fig.~\ref{fig:AB_snapshots}, demonstrating its typical saturation state.

\begin{table*}
	\centering
	\caption{Time-averaged quantities at saturation.
	The columns are (1) case name, (2) dimensionless radial pressure gradient, (3) gas density dispersion$\anote$, (4)--(6) components of the gas velocity dispersion$\bnote$, (7) dust density dispersion$\cnote$, (8)--(10) components of the dust velocity dispersion$\dnote$, and (11) maximum dust density.
	Densities and velocities are in units of the mean gas density $\rhogn$ and the speed of sound $\cs$, respectively.
	In columns~(3) through (10), we show in parentheses the $1\sigma$ time variability of the least significant digit, while in column~(11), we show the $1\sigma$ time variability taken in logarithmic space, in the positive and negative superscripts and subscripts, respectively.
    For Cases~AB and BA, we average all quantities from $t = 5T$ to $\tlim = 10 T$ and from $t = 150T$ to $\tlim = 200T$, respectively.}
	\label{tab:time_averages}
    \begin{tabular}{clddddddddd}
    \hline
    Case & 
    \multicolumn{1}{c}{$\Pi$}            & 
    \multicolumn{1}{c}{$\sigma_{\rhog}$} & 
    \multicolumn{1}{c}{$\sigma_{u_x}$}   & 
    \multicolumn{1}{c}{$\sigma_{u_y}$}   & 
    \multicolumn{1}{c}{$\sigma_{u_z}$}   & 
    \multicolumn{1}{c}{$\sigma_{\rhop}$} & 
    \multicolumn{1}{c}{$\sigma_{v_x}$}   & 
    \multicolumn{1}{c}{$\sigma_{v_y}$}   & 
    \multicolumn{1}{c}{$\sigma_{v_z}$}   & 
    \multicolumn{1}{c}{$\max(\rhop)$}   \\ 
    
    & & 
    \multicolumn{1}{c}{$(10^{-4}\rhogn)$} & 
    \multicolumn{1}{c}{$(10^{-3}\cs)$}    & 
    \multicolumn{1}{c}{$(10^{-3}\cs)$}    & 
    \multicolumn{1}{c}{$(10^{-3}\cs)$}    & 
    \multicolumn{1}{c}{$(\rhogn)$}        & 
    \multicolumn{1}{c}{$(10^{-3}\cs)$}    & 
    \multicolumn{1}{c}{$(10^{-3}\cs)$}    & 
    \multicolumn{1}{c}{$(10^{-3}\cs)$}    & 
    \multicolumn{1}{c}{$(\rhogn)$}       \\ 
         
    \multicolumn{1}{c}{(1)}  &
    \multicolumn{1}{c}{(2)}  &
    \multicolumn{1}{c}{(3)}  &
    \multicolumn{1}{c}{(4)}  &
    \multicolumn{1}{c}{(5)}  &
    \multicolumn{1}{c}{(6)}  &
    \multicolumn{1}{c}{(7)}  &
    \multicolumn{1}{c}{(8)}  &
    \multicolumn{1}{c}{(9)}  &
    \multicolumn{1}{c}{(10)} &
    \multicolumn{1}{c}{(11)} \\ 
    
    \hline       
    \multirow{4}{*}{AB} & 0.01 & 0.074(3) &  2.27(2) &  1.38(3) &  1.67(2) & 1.96(1)  &  1.84(2) &  1.30(2) &  1.52(1) &  64.6^{+  9.7}_{-  8.4} \\
                        & 0.02 & 0.231(7) &  4.37(4) &  2.25(5) &  3.08(4) & 1.934(6) &  3.51(4) &  2.11(4) &  2.74(4) &  61.3^{+  8.8}_{-  7.7} \\
                        & 0.05 & 1.28(6)  & 10.7(3)  &  5.1(3)  &  7.6(3)  & 1.89(2)  &  8.6(2)  &  4.8(3)  &  6.7(3)  &  56.0^{+  8.3}_{-  7.3} \\
                        & 0.1  & 4.9(5)   & 21.(1)   & 10.0(9)  & 15.0(9)  & 1.88(4)  & 17.(1)   &  9.3(9)  & 13.2(8)  &  53.4^{+  8.0}_{-  7.0} \\
    \hline
    \multirow{4}{*}{BA} & 0.01 &  8.(2)   &  2.13(5) &  5.2(1)  & 11.7(2)  & 1.7(2)   &  3.6(1)  &  1.91(6) & 14.6(6)  & 708.5^{+281.4}_{-201.4} \\
                        & 0.02 & 17.(2)   &  4.0(1)  &  9.6(3)  & 26.(1)   & 1.2(2)   &  7.4(3)  &  3.8(2)  & 37.(2)   & 511.2^{+277.7}_{-179.9} \\
                        & 0.05 & 38.(5)   & 10.(1)   & 17.(1)   & 39.(1)   & 0.7(2)   & 17.(1)   &  8.4(8)  & 50.(4)   & 192.1^{+179.8}_{- 92.9} \\
                        & 0.1  & 56.(3)   & 25.(1)   & 28.8(5)  & 63.(1)   & 0.39(2)  & 34.(3)   & 17.(2)   & 63.3(9)  &  24.7^{+  5.7}_{-  4.6} \\
    \hline
    \multicolumn{4}{l}{$\anote$ Defined by equation~\eqref{eq:gas_density_dispersion}}\\
    \multicolumn{4}{l}{$\bnote$ Defined by equation~\eqref{eq:gas_velocity_dispersion}}\\
    \multicolumn{4}{l}{$\cnote$ Analogously defined by equation~\eqref{eq:gas_density_dispersion}}\\
    \multicolumn{4}{l}{$\dnote$ Analogously defined by equation~\eqref{eq:gas_velocity_dispersion}}\\
    \end{tabular}
\end{table*}

\begin{figure*}
	\includegraphics[width=\textwidth]{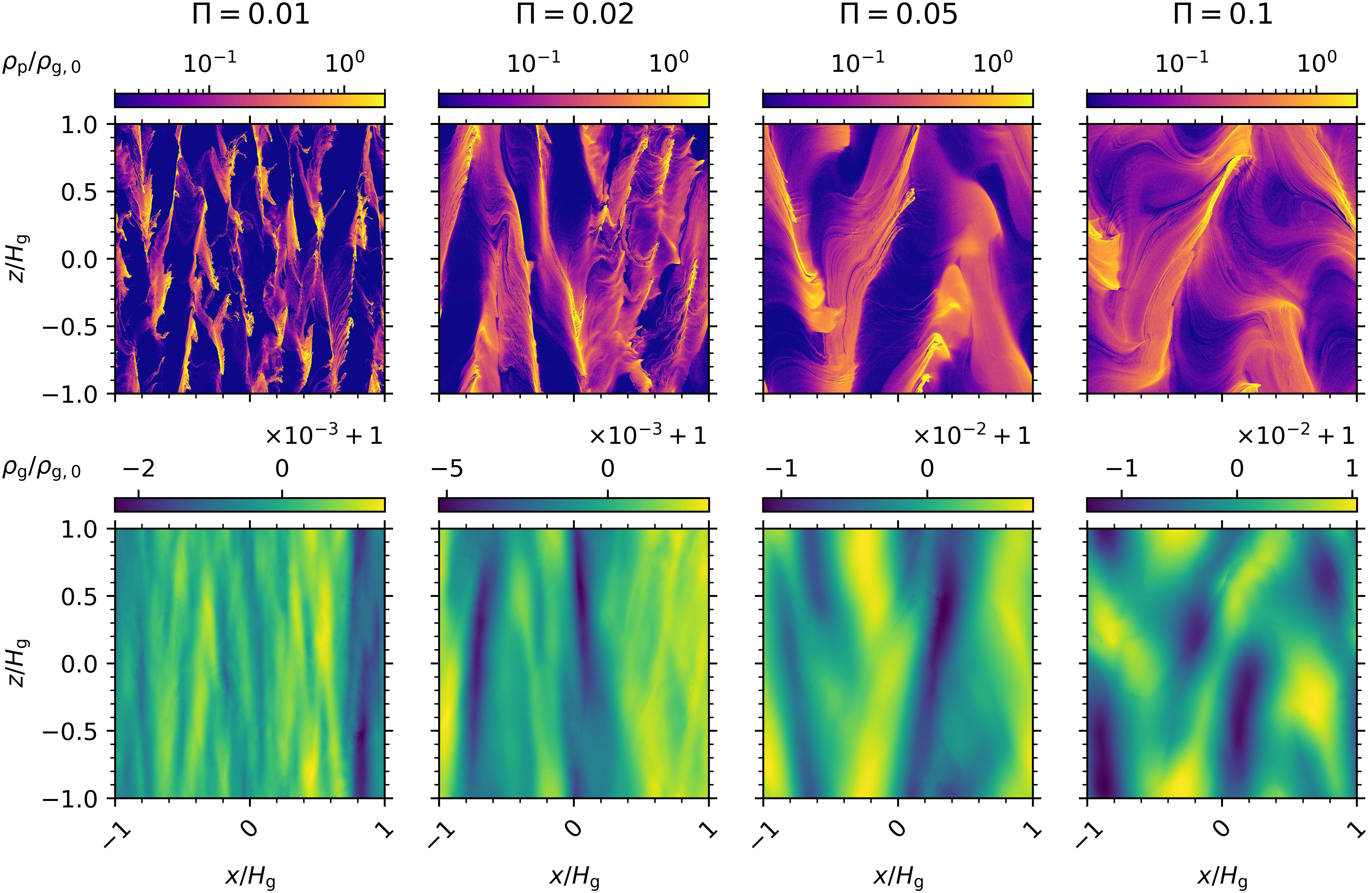}
    \caption{Similar to Fig.~\ref{fig:AB_snapshots}, except for Case~BA ($\taus = 1.0$ with $\epsilon = 0.2$) at $\tlim = 200T$.}
    \label{fig:BA_snapshots}
\end{figure*}

The diagnostics for the BA~models evolve somewhat differently than those for AB.
The density and velocity dispersions do not directly reach a quasi-steady level but instead show roughly three stages of evolution, the morphologies of which can be seen in the available videos (see `Data Availability' section).
In the initial stage, as shown in Fig.~\ref{fig:dispersions}, the dispersions first increase rapidly until about $10T$ to $20T$.
The end of the first stage is roughly delineated by the knees of $\sigma_{\rhop}$, and this corresponds to the time when the initial regular pattern of short slanted dust filaments begins to roll and break.
In the second stage, the dispersions continue to increase at a much slower rate, for instance in $\sigma_{\rhog}$ and $\sigma_{u_z,v_z}$ for $\Pi = 0.01$ between $15T$ and $100T$, during which dust filaments are in the process of merging with each other.
By the third stage, the dispersions finally reach a quasi-steady level, when it appears that no further major mergers occur.
We note that models with higher values of $\Pi$ reach saturation earlier than models with lower values.
\citet[][\S~3.2 and fig.~2]{JohansenYoudin2007} describe the early growth stage of their Run~BA for $\Pi = 0.05$.
Fig.~\ref{fig:BA_snapshots} shows the final snapshots at $\tlim = 200T$ of the dust and gas density fields during saturation for Case~BA where we note the prominence of long, vertically-slanted, dense dust filaments \citep{JohansenYoudin2007, YangZhu2021}.
Judging from Fig.~\ref{fig:dispersions}, we average all reported quantities for the saturation state of all BA~models from $t = 150T$ to $\tlim = 200T$ throughout this work.

Time-averaged dispersions for BA~models during saturation are listed in Table~\ref{tab:time_averages}.
As shown in the table, $\sigma_{\rhop}$ decreases as $\Pi$ increases, differing by more than a factor of four between $\Pi = 0.01$ and $0.1$ models.
Conversely, $\sigma_{\rhog}$ increases as $\Pi$ increases,  differing by almost an order of magnitude between $\Pi = 0.01$ and $0.1$ models.
However, a maximum density fluctuation of order $10^{-3}\rhogn$ in the latter indicates the gas remains relatively incompressible overall.
As for $\sigma_{\uvec}$ and $\sigma_{\vvec}$, we find both increase as $\Pi$ increases.
Comparing the dust and the gas for any given $\Pi$, $\sigma_{v_x} > \sigma_{u_x}$ but $\sigma_{u_y} > \sigma_{v_y}$.
Lastly, comparing the components in general, $\sigma_{u_z} > \sigma_{u_y} > \sigma_{u_x}$ for the gas, while $\sigma_{v_z} > \sigma_{v_x} > \sigma_{v_y}$ for the dust, as was similarly found for multi-species Model~B by \cite{YangZhu2021}.

\subsection{Morphology}
\label{sec:morphology}

In this section, we study the morphology in the density fields of our models at non-linear saturation.
Specifically, we examine snapshots of the dust and gas density fields, the spatial correlation of these snapshots, and the distributions of these densities.
We start with Case~BA, where $\taus = 1.0$ for marginally coupled particles, in Section~\ref{sec:BA_morphology} followed by Case~AB, where $\taus = 0.1$ for tightly coupled particles, in Section~\ref{sec:AB_morphology}.

\subsubsection{Marginally-coupled particles}
\label{sec:BA_morphology}

Final snapshots of the dust and gas density fields for each BA~model at saturation are shown in Fig.~\ref{fig:BA_snapshots}.
As originally reported by \cite{JohansenYoudin2007}, the dust shows a sharp pattern of dense filaments aligned vertically with slight tilts in alternating directions, extending to about the height of our domain of $2\Hg\times2\Hg$.
As the videos show (see `Data Availability' section), these filaments act like traffic jams that impede and collect any particles drifting radially between them, while each filamentary pattern coherently moves either upward or downward \citep{JohansenYoudin2007, YangJohansen2014, YangMacLowJohansen2018}.
Comparing the dust and gas snapshots for each model, we find the gas also condenses into vertical structures somewhat in between the dust filaments while remaining relatively incompressible overall \citep{YangJohansen2014, LiYoudinSimon2018, YangZhu2021}.

As shown in Fig.~\ref{fig:BA_snapshots}, the morphology of Case~BA changes with $\Pi$.
As $\Pi$ decreases, the radial separation between dense dust filaments decreases, and we find fewer particles drifting in between.
Dust filaments also show more vertical segmentation and steeper tilts.
The gas also shows fluctuations of shorter radial characteristic length.
For $\Pi = 0.01$, there appears to be a large-scale radial variation in gas density on the order of $L_x$.

\begin{figure*}
	\includegraphics[width=\textwidth]{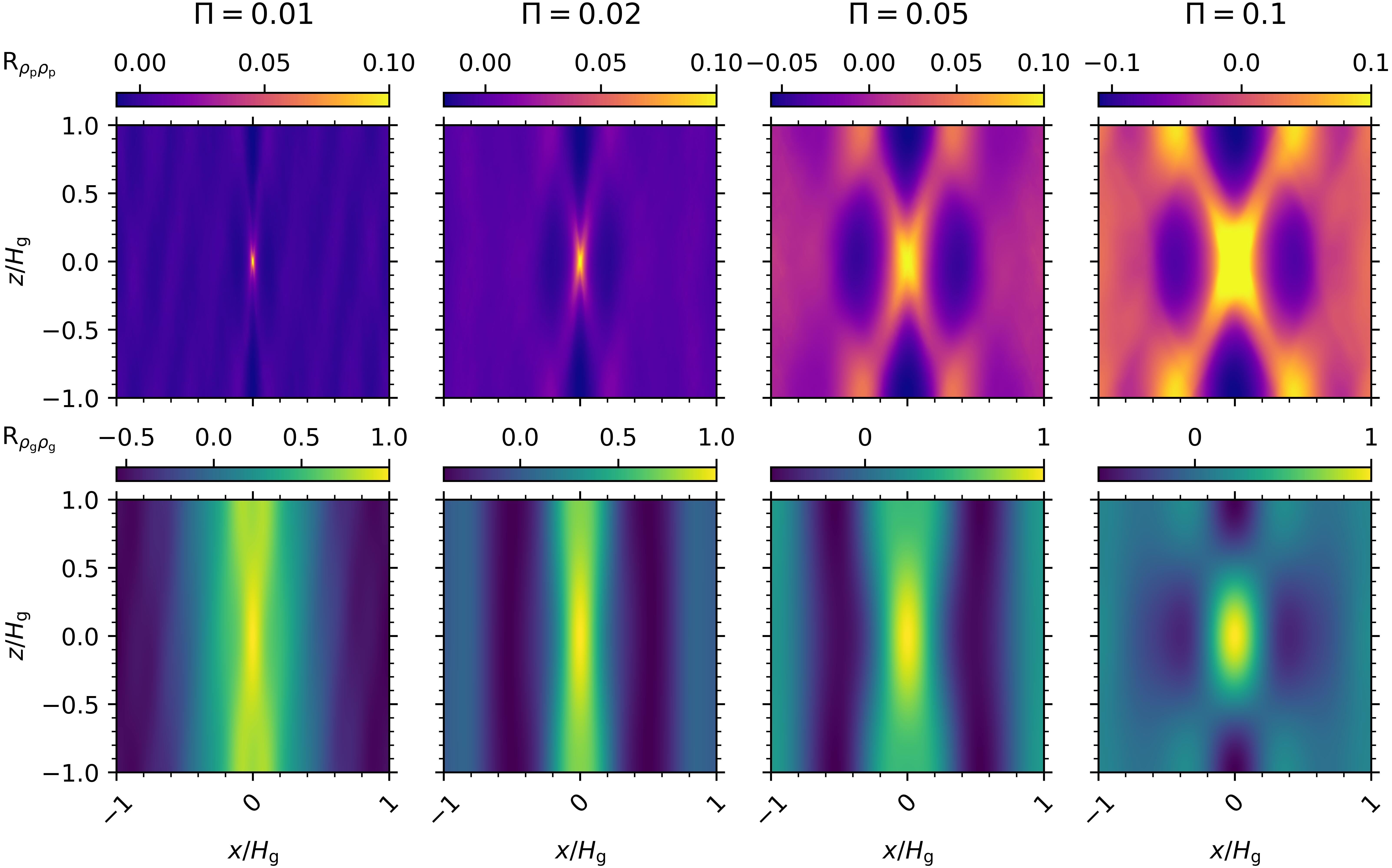}
    \caption{Time-averaged normalised spatial auto-correlation functions for the dust (top) and the gas (bottom) density fields, defined by equation~\eqref{eq:auto-correlation}, for Case~BA ($\taus = 1.0$ with $\epsilon = 0.2$)
    The notations $x$ and $z$ in this context are the radial and vertical displacements, respectively.
    The columns from left to right show models with increasing radial pressure gradient $\Pi$.}
    \label{fig:BA_avgRs}
\end{figure*}

In order to quantify the structural dependence on $\Pi$, we compute the spatial correlation (a.k.a. structure function) of the dust density field.
It is quantified by the normalised auto-correlation of a density field $\rho$:
\begin{equation}
    \mathrm{R}_{\rho\rho}=
    \frac{\mathcal{F}^{-1}\left(\hat{\delta\rho}\hat{\delta\rho}^*\right)}
         {L_x L_y L_z\langle\rho^2\rangle},
	\label{eq:auto-correlation}
\end{equation}
where $\mathcal{F}^{-1}$ is the inverse Fourier transform operator, $\hat{\delta\rho}$ is the Fourier transform of the local density deviation $\delta\rho = \rho - \langle\rho\rangle$, $\hat{\delta\rho}^*$ is the complex conjugate of the transform, and the notation $\langle \cdot \rangle$ is the volume average defined by equation~\eqref{eq:vol-avg}.
The top row of Fig.~\ref{fig:BA_avgRs} shows the time-averaged spatial correlation of the dust density field $\mathrm{R}_{\rhop\rhop}$ for each BA~model, where $x$ and $z$ in this context are the radial and vertical displacements, respectively.
The signals shown by the spatial correlation reveal repeating patterns in the dust density field, which in this case align with the upward and downward floating patterns of dust filaments.
Similar to the unit cell used to describe a crystal structure, each signal contains a characteristic bright cross that is tiled.
As $\Pi$ decreases, we find more adjacent tiles in the radial direction, which correspond to the decrease in radial separation between dense filaments.
The signal of the central cross becomes less vertically extended, consistent with more vertical segmentation of filaments.
Fig.~\ref{fig:BA_avgRs} also indicates the absolute slopes of the crosses increase as $\Pi$ decreases, corresponding to the steeper tilts of filaments.
Similarly, we compute the time-averaged spatial correlation $\mathrm{R}_{\rhog\rhog}$ of the gas density fields, which is shown in the bottom row of Fig.~\ref{fig:BA_avgRs}.
From $\Pi = 0.01$ to $0.05$, we see signals of pure radial modes of decreasing length from roughly $L_x$ to $L_x/2$, which supports the large-scale radial variation seen in Fig.~\ref{fig:BA_snapshots} for $\Pi = 0.01$.

\begin{figure*}
	\includegraphics[width=\textwidth]{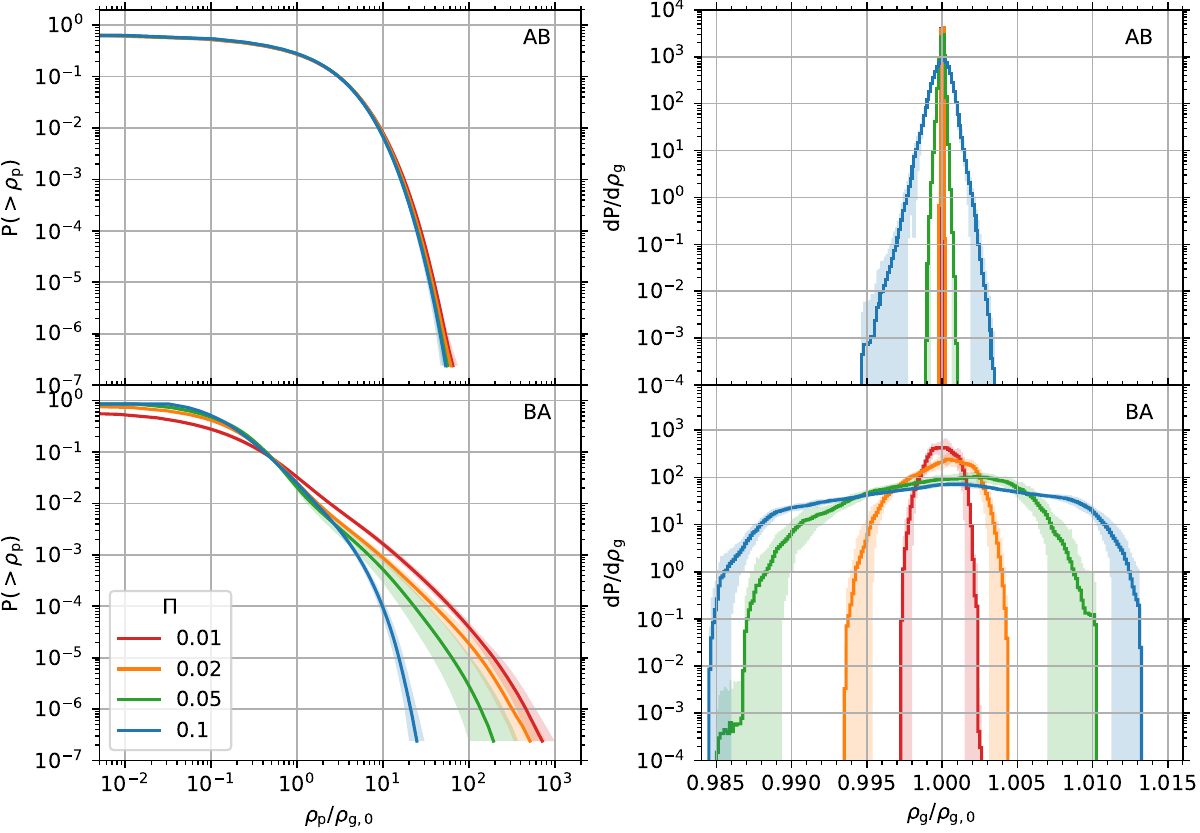}
    \caption{Time-averaged cumulative distribution functions for the dust density (left column) and probability density functions for the gas density (right column) for each model.
    The top and bottom rows are for Cases~AB ($\taus = 0.1$ with $\epsilon = 1.0$) and BA ($\taus = 1.0$ with $\epsilon = 0.2$), respectively.
    Solid lines represent the time-averaged densities, shaded areas represent the $1\sigma$ time variability, and different colours represent models with different values of $\Pi$.}
    \label{fig:densities}
\end{figure*}

Lastly, we find for Case~BA the cumulative distribution function $\rhopCDF$ for the dust density and the probability density function $\dPdrhog$ for the gas density.
We time average both distributions and show the results and variabilities in Fig.~\ref{fig:densities}.
The maximum dust densities $\max(\rhop)$, shown in the right-hand tail of $\rhopCDF$ and listed in Table~\ref{tab:time_averages}, increase with decreasing $\Pi$, differing by more than one order of magnitude between $\Pi=0.1$ and $0.01$.
For $\rhopCDF$ of all BA~models in Fig.~\ref{fig:densities}, we note approximately 8\% of the computational domain has $\rhop \approx 0.5\rhogn$.
As $\Pi$ increases, $\rhopCDF$ becomes steeper, indicating the dust is more evenly distributed throughout the domain, which is consistent with more particles shown streaming between dense dust filaments (Fig.~\ref{fig:BA_snapshots}).
For the gas, $\dPdrhog$ also varies between BA~models, becoming less Gaussian and more negatively skewed as $\Pi$ increases.
Moreover, the width of the distribution increases as $\Pi$ increases, differing by an order of magnitude between $\Pi = 0.01$ and 0.1.
Lastly, we note the average logarithmic slope of $\dPdrhog$ changes a few times about the mean at $1\rhogn$.

\subsubsection{Tightly-coupled particles}
\label{sec:AB_morphology}

Fig.~\ref{fig:AB_snapshots} shows the final snapshots of the dust and gas density fields for AB~models, where the particles are more tightly coupled to the gas.
At saturation, each system forms a collection of turbulent vortices of various sizes, each surrounded by filamentary structures of dust, as first shown by \citet[fig.~5]{JohansenYoudin2007} and as similarly described for the multi-species Model~Af by \cite{YangZhu2021}.
More importantly, the morphology of Case~AB changes with $\Pi$.
The turbulent vortices appear to become larger as $\Pi$ increases, and we attempt to quantify this effect below.

\begin{figure}
	\includegraphics[width=\columnwidth]{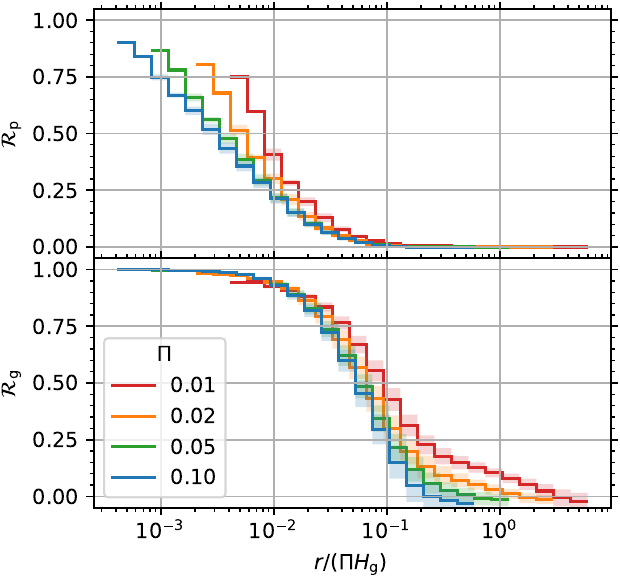}
    \caption{Time-averaged radial profiles (Section~\ref{sec:AB_morphology}) of the normalised spatial auto-correlation functions for the dust (top) and the gas (bottom) density fields, defined by equation~\eqref{eq:auto-correlation}, for Case~AB ($\taus = 0.1$ with $\epsilon = 1.0$).
    The notation $r$ in this context is the radial displacement.
    Solid lines represent the average values of each radial bin over time, shaded areas represent the $1\sigma$ time variability, and different colours represent models with different values of $\Pi$.
    Radial bins are scaled by $\Pi$.}
    \label{fig:AB_avgRs_rad-prof}
\end{figure}

As in Section~\ref{sec:BA_morphology}, we compute the normalised spatial auto-correlation function of the dust density field via equation~\eqref{eq:auto-correlation}.
Because $\mathrm{R}_{\rhop\rhop}$ of each AB~model appears isotropic around a central peak, we approximate the average radial profile $\Rp(r)$ by binning the discrete 2D data in $r$, where $r$ is the radial displacement in this context.
The top panel of Fig.~\ref{fig:AB_avgRs_rad-prof} compares the time-averaged radial profiles for the dust $\Rp(r)$ between models with different $\Pi$, where we note $r$ is normalised by $\Pi\Hg$.
We find the width of the profile roughly scales with $\Pi$, indicating increasing sizes of the dust--gas vortices, as observed in Fig.~\ref{fig:AB_snapshots}.
However, for models with smaller $\Pi$, the width of the profile is slightly wider than a simple linear relationship with $\Pi$ would infer.
On the other hand, the length scales (or the velocity scales) in the linear system for the streaming instability appear to be linearly proportional to $\Pi$ \citep{YoudinGoodman2005, YoudinJohansen2007}.
Therefore, non-linear effect(s) of the streaming instability may contribute to the deviation seen here from what a linear system would predict.

Similarly, we compute the time-averaged radial profile $\Rg(r)$ of the gas density spatial correlation $\mathrm{R}_{\rhog\rhog}$.
Comparing half widths at half maximum, we note the $\mathrm{R}_{\rhog\rhog}(x = 0)$ in the vertical direction are slightly wider than $\mathrm{R}_{\rhog\rhog}(z = 0)$ in the radial direction -- almost within a $1\sigma$ time variability for $\Pi = 0.1$ but up to a 20\% difference for $\Pi = 0.01$ -- which we do not find for the dust.
Assuming symmetry, the bottom panel of Fig.~\ref{fig:AB_avgRs_rad-prof} compares the resulting radial profiles $\Rg(r)$ between AB~models.
The width of the profile approximately scales with $\Pi$, consistent with larger dust--gas vortices seen in Fig.~\ref{fig:AB_snapshots}, but is wider for smaller $\Pi$, similar to what is shown in the dust component.

As in Section~\ref{sec:BA_morphology}, we find the cumulative distribution function $\rhopCDF$ for the dust density and the probability density function $\dPdrhog$ for the gas density for Case~AB.
In Fig.~\ref{fig:densities}, we plot the time averages and variabilities of these distributions.
For the dust, we find $\rhopCDF$ barely changes with $\Pi$.
As shown in Table~\ref{tab:time_averages}, the maximum dust density $\max(\rhop)$ of AB~models decreases as $\Pi$ increases but shows a relative difference of at most 19\% between $\Pi = 0.01$ and $0.1$.
This small change in $\rhopCDF$ and $\max(\rhop)$ with $\Pi$ is consistent with the relative insensitivity of the dust density dispersion at saturation to changes in $\Pi$ for Case~AB (Section~\ref{sec:saturation_state} and Fig.~\ref{fig:dispersions}).
As for the gas, $\dPdrhog$ appears mostly Gaussian albeit somewhat negatively skewed.
As $\Pi$ increases, the width of the distribution increases, differing by more than one order of magnitude between $\Pi = 0.01$ and 0.1.

\subsection{Kinematics}
\label{sec:kinematics}

In this section, we study the kinematics of the gas and dust in our models at non-linear saturation.
We first examine gas turbulence in Subsection~\ref{sec:gas_turbulence} by measuring the velocity distribution, Mach number, and $\alpha$ parameter.
Then we analyse dust motions in Subsection~\ref{sec:dust_motions} by finding the velocity distribution and diffusion coefficient of the particles and by estimating the dust scale height.

\subsubsection{Gas turbulence}
\label{sec:gas_turbulence}

\begin{figure*}
	\includegraphics[width=\textwidth]{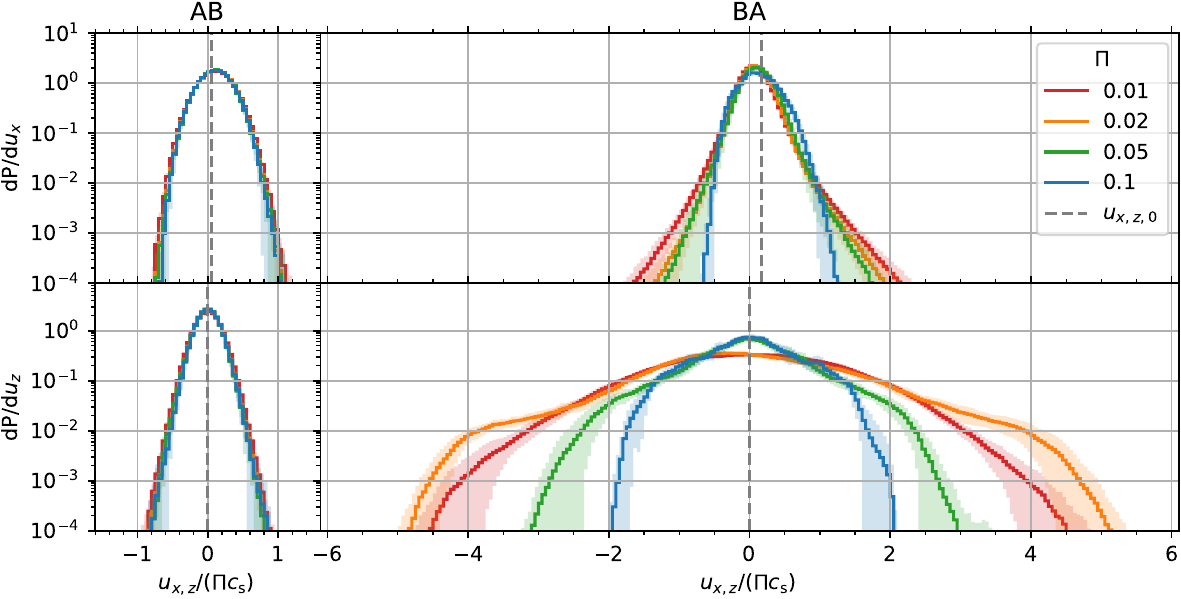}
    \caption{Time-averaged probability density functions for the radial (top row) and vertical (bottom row) components of the gas velocity for each model.
    The left and right columns are for Cases~AB ($\taus = 0.1$ with $\epsilon = 1.0$) and BA ($\taus = 1.0$ with $\epsilon = 0.2$), respectively.
    Solid lines represent the time-averaged distributions, and shaded areas represent the $1\sigma$ time variability.
    Different colours represent models with different values of $\Pi$, and the dashed grey vertical lines represent components of the initial equilibrium velocity $u_{x,z,0}$.
    Velocity bins are scaled by $\Pi$.}
    \label{fig:dPduxz}
\end{figure*}

For each of our models, we find the probability density function $\dPduxz$ for the radial and vertical components of the gas velocity.
We time average the distributions and show the results along with variabilities in Fig.~\ref{fig:dPduxz}, where we note $u_{x,z}$ is normalised by $\Pi\cs$.
For AB~models, each component of $\dPduxz$ appears Gaussian, and the time variability in each velocity bin seems to increase as $\Pi$ increases.
When normalised by $\Pi$, $\dPduxz$ is nearly coincident as also evident by the average $\avgux$ and standard deviations $\sigma_{u_{x,z}}$ listed in Table~\ref{tab:velocities}, indicating the gas velocity linearly scales with $\Pi$ in this case.
Moreover, while $\dPduz$ is symmetrical about $\avguz \approx u_{z,0} = 0$, more of the gas migrates radially outward at saturation than at the initial equilibrium with $\avgux > u_{x,0} \approx 0.05\Pi\cs$.

\begin{table*}
	\centering
	\caption{Initial equilibrium gas and dust velocities and the averages and standard deviations at saturation.
	The columns are (1) case name, (2) dimensionless radial pressure gradient, radial component of the (3) initial equilibrium (4) average, and (5) standard deviation of the gas velocity; (6) standard deviation of the vertical component of the gas velocity; and (7)--(10) the same as (3)--(6) but for the dust.
	Velocities are normalised to $\Pi\cs$.}
	\label{tab:velocities}
    \begin{tabular}{clcrrrcrrr}
    \hline
    Case                               &  
    \multicolumn{1}{c}{$\Pi$}          &  
    \multicolumn{1}{c}{$u_{x,0}$}      &  
    \multicolumn{1}{c}{$\avgux$}       &  
    \multicolumn{1}{c}{$\sigma_{u_x}$} &  
    \multicolumn{1}{c}{$\sigma_{u_z}$} &  
    \multicolumn{1}{c}{$v_{x,0}$}      &  
    \multicolumn{1}{c}{$\avgvx$}       &  
    \multicolumn{1}{c}{$\sigma_{v_x}$} &  
    \multicolumn{1}{c}{$\sigma_{v_z}$} \\ 
    
    & &
    \multicolumn{1}{c}{($\Pi\cs$)} &  
    \multicolumn{1}{c}{($\Pi\cs)$} &  
    \multicolumn{1}{c}{($\Pi\cs)$} &  
    \multicolumn{1}{c}{($\Pi\cs)$} &  
    \multicolumn{1}{c}{($\Pi\cs)$} &  
    \multicolumn{1}{c}{($\Pi\cs)$} &  
    \multicolumn{1}{c}{($\Pi\cs)$} &  
    \multicolumn{1}{c}{($\Pi\cs)$} \\ 
    
    \multicolumn{1}{c}{(1)}  &
    \multicolumn{1}{c}{(2)}  &
    \multicolumn{1}{c}{(3)}  &
    \multicolumn{1}{c}{(4)}  &
    \multicolumn{1}{c}{(5)}  &
    \multicolumn{1}{c}{(6)}  &
    \multicolumn{1}{c}{(7)}  &
    \multicolumn{1}{c}{(8)}  &
    \multicolumn{1}{c}{(9)}  &
    \multicolumn{1}{c}{(10)} \\ 
    
    \hline       
    \multirow{4}{*}{AB} & 0.01 & \multirow{4}{*}{0.05} & 0.11 & 0.23 & 0.17 & \multirow{4}{*}{-0.05} & -0.11 & 0.18 & 0.15 \\
                        & 0.02 &                       & 0.11 & 0.22 & 0.15 &                        & -0.11 & 0.18 & 0.14 \\
                        & 0.05 &                       & 0.11 & 0.21 & 0.15 &                        & -0.11 & 0.17 & 0.13 \\
                        & 0.1  &                       & 0.11 & 0.22 & 0.15 &                        & -0.11 & 0.17 & 0.13 \\
    \hline
    \multirow{4}{*}{BA} & 0.01 & \multirow{4}{*}{0.16} & 0.07 & 0.21 & 1.17 & \multirow{4}{*}{-0.82} & -0.33 & 0.36 & 1.47 \\
                        & 0.02 &                       & 0.09 & 0.20 & 1.29 &                        & -0.47 & 0.37 & 1.84 \\
                        & 0.05 &                       & 0.11 & 0.21 & 0.77 &                        & -0.57 & 0.35 & 1.01 \\
                        & 0.1  &                       & 0.13 & 0.25 & 0.63 &                        & -0.64 & 0.35 & 0.63 \\
    \hline
    \end{tabular}
\end{table*}

The gas velocity distributions for Case~BA in Fig.~\ref{fig:dPduxz}, on the other hand, show more variation between models with different $\Pi$.
While $\avguz \approx u_{z,0} = 0$ for all $\Pi$, $\avgux / \Pi$ increases by about a factor of two from $\Pi = 0.01$ to 0.1, as shown in Table~\ref{tab:velocities}.
Moreover, $\sigma_{u_z} / \Pi$ slightly increases from $\Pi = 0.01$ to 0.02 but monotonically decreases from 0.02 to 0.1, while $\sigma_{u_x} / \Pi$ are roughly equal for all BA~models.
At saturation, $\avgux < u_{x,0} \approx 0.16\Pi\cs$ as more of the gas shifts towards the negative side of the initial equilibrium.
For each BA~model, $\dPdux$ skews somewhat positively and becomes less Gaussian as $\Pi$ decreases.
On the other hand, the probability $\dPduz$ appears symmetrical but significantly wider; however, the width has no obvious trend with $\Pi$.

\begin{figure}
	\includegraphics[width=\columnwidth]{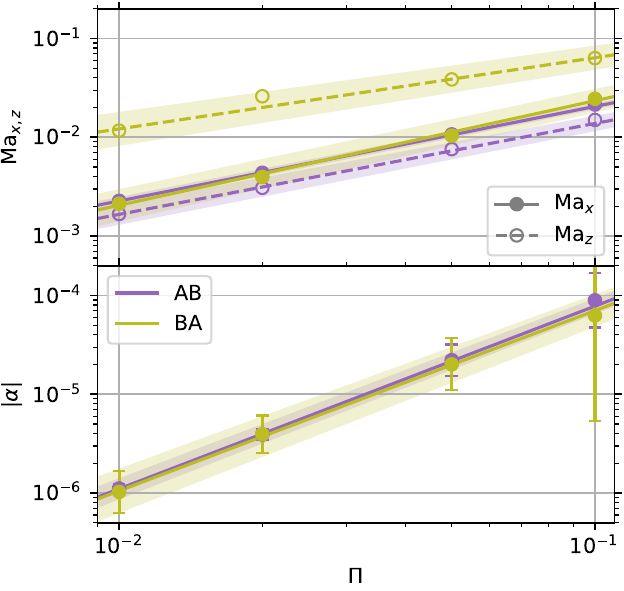}
    \caption{Time-averaged Mach numbers (top) and $\alpha$ parameters (bottom) of the gas as a function of $\Pi$ for Cases~AB ($\taus = 0.1$ with $\epsilon = 1.0$) and BA ($\taus = 1.0$ with $\epsilon = 0.2$).
    In both panels, different colours represent different cases, lines represent the best-fitting power-law defined by equation~\eqref{eq:power_law}, and shaded areas represent the $1\sigma$ relative uncertainty in fitting parameters.
    In the top panel, we do not show time variabilities as they are too small to be seen, solid lines and filled circles correspond to the radial component, while dashed lines and open circles correspond to the vertical component.
    In the bottom panel, we plot the absolute value of $\alpha$, where $\alpha < 0$ for Case~AB and $\alpha > 0$ for BA, and show a $3\sigma$ time variability for exaggeration purposes.}
    \label{fig:gas_turbulence}
\end{figure}

Next, we compute the characteristic Mach number of the gas motions for each model in the radial and vertical directions as $\Ma_{x,z} = \sigma_{u_{x,z}} / \cs$, where $\sigma_{u_{x,z}}$ is the gas velocity dispersion defined by equation~\eqref{eq:gas_velocity_dispersion}.
The top panel of Fig.~\ref{fig:gas_turbulence} shows the time-averaged $\Ma$ as a function of $\Pi$.
We note our results for $\Ma(\Pi = 0.05)$ differ at most by 14\% from those of \citet[table~2]{JohansenYoudin2007}.
We find in general $\Ma$ increases as $\Pi$ increases.
For Case~AB, $\Ma_x \gtrsim \Ma_z$, as similarly described for the multi-species Model~Af by \cite{YangZhu2021}, and by about 40\% on average in our case.
Meanwhile for Case~BA, $\Ma_z > \Ma_x$, the same relationship found for multi-species Model~B by \cite{YangZhu2021}, and by a factor of 3--7 for all our BA~models.
The significant anisotropy in the gas velocity dispersion for Case~BA is consistent with the different widths of the gas velocity distributions between radial and vertical components (Fig.~\ref{fig:dPduxz}).
Notably, $\Ma_x(\Pi)$ is roughly equal between Cases~AB and BA, while $\Ma_z(\Pi)$ for Case~BA is significantly larger than that for AB.

We fit a power law
\begin{equation}
    f(\Pi) = a\Pi^k.
	\label{eq:power_law}
\end{equation}
to each Ma, using the $1\sigma$ time variability as the uncertainty.
The best-fitting lines and parameters are shown in Fig.~\ref{fig:gas_turbulence} and listed in Table~\ref{tab:gas_fitting_parameters}, respectively.
For both components of Case~AB and $\Ma_x(\Pi)$ of BA, the power-law index $k$ is approximately unity, indicating these gas velocity dispersions linearly scale with $\Pi$.
However, $\Ma_z$ of Case~BA does not appear to linearly scale with $\Pi$ as $k \approx 3/4$.

\begin{table*}
	\centering
	\caption{Parameters for the gas of the best-fitting power laws$\anote$ shown in Fig.~\ref{fig:gas_turbulence}.
	The columns correspond to (1) case name, (2)--(3) parameters for the radial component of the gas Mach number, (4)--(5) for the vertical component of the gas Mach number, and (6)--(7) for the $\alpha$ parameter of the gas.}
	\label{tab:gas_fitting_parameters}
    \begin{tabular}{cllllrl}
    \hline
    Case & 
    \multicolumn{2}{c}{$\Ma_x(\Pi)$}  &     
    \multicolumn{2}{c}{$\Ma_z(\Pi)$}  &     
    \multicolumn{2}{c}{$\alpha(\Pi)$} \\    
    
    &
    \multicolumn{1}{c}{$a$} &               
    \multicolumn{1}{c}{$k$} &               
    \multicolumn{1}{c}{$a$} &               
    \multicolumn{1}{c}{$k$} &               
    \multicolumn{1}{c}{$a/10^{-3}$} &  
    \multicolumn{1}{c}{$k$} \\              
         
    \multicolumn{1}{c}{(1)}  &
    \multicolumn{1}{c}{(2)}  &
    \multicolumn{1}{c}{(3)}  &
    \multicolumn{1}{c}{(4)}  &
    \multicolumn{1}{c}{(5)}  &
    \multicolumn{1}{c}{(6)}  &
    \multicolumn{1}{c}{(7)}  \\
    
    \hline       
    \multirow{1}{*}{AB} & $0.19 \pm 5\%$  & $0.96 \pm 1\%$ & $0.11 \pm 12\%$ & $0.92 \pm 3\%$ & $-5.53 \pm 12\%$ & $1.85 \pm 2\%$ \\
    
    \multirow{1}{*}{BA} & $0.27 \pm 17\%$ & $1.06 \pm 4\%$ & $0.34 \pm 17\%$ & $0.72 \pm 7\%$ &  $4.70 \pm 24\%$ & $1.82 \pm 3\%$ \\
    \hline
    \multicolumn{3}{l}{$\anote$ Defined by equation~\eqref{eq:power_law}} \\
    \end{tabular}
\end{table*}

Another important property is the shear stress that drives gas accretion. 
For this reason, we compute the $\alpha$ parameter as $\alpha = W_{xy} / \cs^2$ \citep[][eq.~48]{ShakuraSunyaev1973, BalbusHawley1998}, where $W_{xy} \equiv \langle\rhog(\delta u_x - \Delta u_x)(\delta u_y - \Delta u_y)\rangle$ is the radial--azimuthal component of the Reynolds stress tensor, in which the notation $\langle \cdot \rangle$ is the volume average defined by equation~\eqref{eq:vol-avg}, $\delta\uvec \equiv \uvec - \uvec_0$ is the local gas velocity deviation from the initial equilibrium velocity $\uvec_0$, and $\Delta\uvec$ is the mass-weighted average gas velocity deviation defined by equation~\eqref{eq:gas-velocity-deviation}.
The stress $W_{xy}$ describes the angular momentum flux, and its sign describes the direction of the flux.\footnote{
It remains unclear if the sign of the stress $W_{xy}$ is indicative of the diffusive property of the gas as suggested by \cite{YangZhu2021}.}
As in the multi-species Models~Af and B by \cite{YangZhu2021}, we find in general $W_{xy} < 0$ and $W_{xy} > 0$ for Cases~AB and BA, respectively.

The bottom panel of Fig.~\ref{fig:gas_turbulence} shows the absolute value of time-averaged $\alpha$ as a function of $\Pi$.
In general, we find $|\alpha|$ increases as $\Pi$ increases and note it is roughly equal between cases.
Since the streaming instability is powered by the relative velocity between the gas and dust, e.g., driven by radial pressure gradients \citep{YoudinGoodman2005}, increasing $\Pi$ injects more energy into the system and increases the turbulence strength, as seen by this trend in $|\alpha|$ and in changes to the dust density fields (Figs.~\ref{fig:AB_snapshots} and \ref{fig:BA_snapshots}).
We fit the power law defined by equation~\eqref{eq:power_law} to each $\alpha$, using the $1\sigma$ time variability as the uncertainty, and show the best-fitting lines and parameters in Fig.~\ref{fig:gas_turbulence} and Table~\ref{tab:gas_fitting_parameters}, respectively.
In the linear system for the streaming instability, velocities appear to linearly scale with $\Pi$ \citep{YoudinGoodman2005, YoudinJohansen2007}.
Thus, as $\alpha \propto W_{xy}$, which in turn is proportional to the product of gas velocity deviations, $\alpha$ may scale by $\Pi^2$ in a linear system from purely dimensional arguments.
However, in both non-linear Cases~AB and BA, we find the power-law index $k \approx 1.8$, which is somewhat less than two.

\subsubsection{Dust motions}
\label{sec:dust_motions}

\begin{figure*}
	\includegraphics[width=\textwidth]{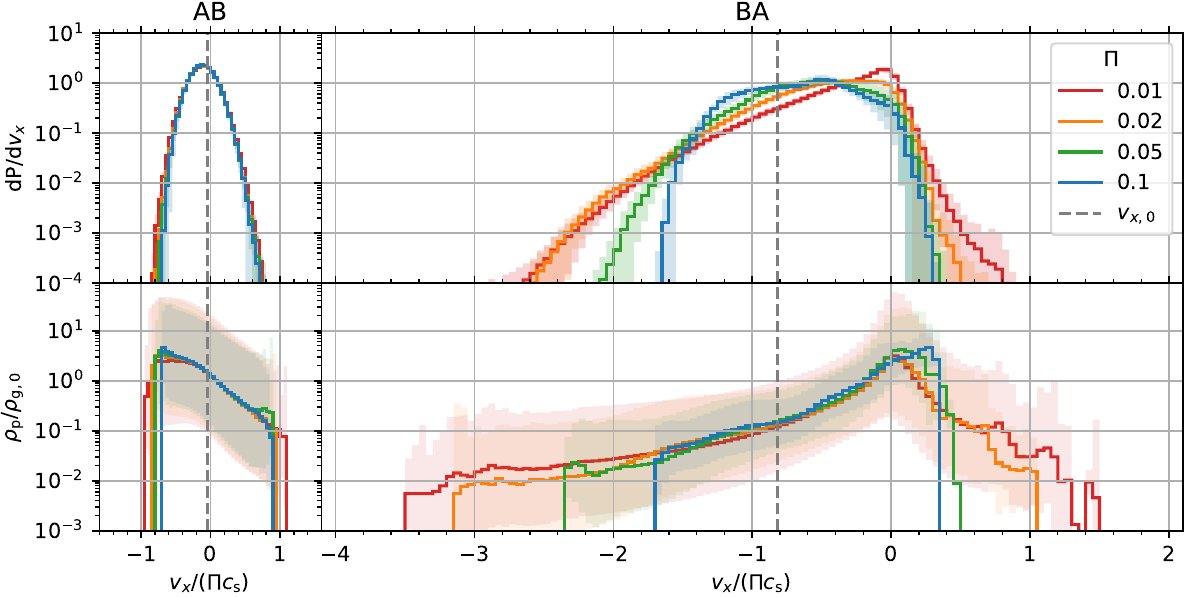}
    \caption{Time-averaged probability of finding a particle with a particular radial velocity (top row) and the average particle density over time and space of cells with similar radial particle velocities (bottom row) for each model.
    Different colours represent models with different values of $\Pi$, shaded areas represent the $1\sigma$ time variability (top row) or the standard deviation taken in logarithmic space (bottom row), and the dashed grey vertical lines represent the radial component of the initial equilibrium velocity $v_{x,0}$.
    Velocities are scaled by $\Pi$.}
    \label{fig:dPdvx}
\end{figure*}

\begin{figure*}
	\includegraphics[width=\textwidth]{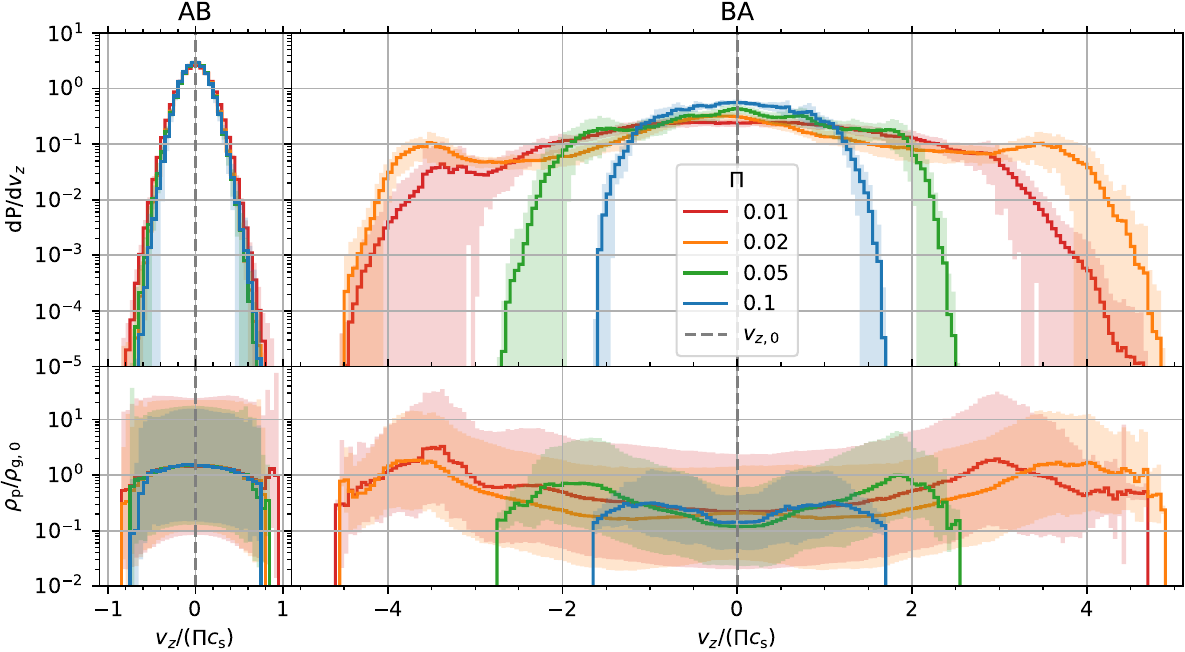}
    \caption{Similar to Fig.~\ref{fig:dPdvx}, except for the vertical component.}
    \label{fig:dPdvz}
\end{figure*}

Similar to what was done for the gas (Section~\ref{sec:gas_turbulence}), we compute the probability $\dPdvxz$ of finding a particle with a particular velocity in each of our models.
The top rows of Figs.~\ref{fig:dPdvx} and \ref{fig:dPdvz} show the radial and vertical components, respectively, of the time-averaged dust velocity distributions, where $v_{x,z}$ is normalised by $\Pi\cs$.
For AB~models, $\dPdvz$ appears Gaussian, while $\dPdvx$ skews somewhat positively.
Moreover, $\dPdvxz$ seems to linearly scale with $\Pi$, as also evident in Table~\ref{tab:velocities}.
As originally reported by \citet[Runs~AB and AC]{JohansenYoudin2007} and by \citet[Model~Af]{YangZhu2021}, more of the dust migrates radially inward at saturation, with $\avgvx < v_{x,0} \approx -0.05\Pi\cs$ in Table~\ref{tab:velocities}.

In addition, we average over time and space the particle density $\rhop$ of cells with similar particle velocities, and the results are shown in the bottom rows of Figs.~\ref{fig:dPdvx} and \ref{fig:dPdvz}.
For each AB~model in Fig.~\ref{fig:dPdvx}, we find the average $\rhop$ within radial velocity bins decreases by more than one order of magnitude as $v_x$ becomes more positive \citep[cf.][fig.~16, Run~AB]{JohansenYoudin2007}. 
This can be seen in the videos for Case~AB (see `Data Availability' section), where inward flowing filaments appear denser while outward flowing filaments appear less dense.
In Fig.~\ref{fig:dPdvz}, the average $\rhop$ within vertical velocity bins is somewhat uniform across the range of $v_z$, but we note for $\Pi = 0.01$ a possibly spurious maxima in the average $\rhop$ in the positive tail for $v_z \approx 0.9\Pi\cs$, which may be due to the small number of cells in the tail.

For Case~BA, on the other hand, the radial and vertical dust velocity distributions of particles in the top rows of Figs.~\ref{fig:dPdvx} and \ref{fig:dPdvz}, respectively, vary more between models with different $\Pi$, as was found for the gas (Fig.~\ref{fig:dPduxz} and Section~\ref{sec:gas_turbulence}).
First, we note that since the total momentum of the gas and dust remains conserved for each AB or BA~model, $\avgux \approx -\epsilon\avgvx$ as shown in Table~\ref{tab:velocities}.
Consequently for Case~BA, as $\avgvx / \Pi$ nearly doubles from $\Pi = 0.01$ to 0.1, so does $\avgux / \Pi$.
Similarly, $\dPdvx$ (Fig.~\ref{fig:dPdvx}) and $\dPdux$ (Fig.~\ref{fig:dPduxz}) skew more negatively and positively, respectively.
Concurrent with the gas in Table~\ref{tab:velocities}, $\sigma_{v_x} / \Pi$ are roughly equal for all BA~models while $\sigma_{v_z} / \Pi$ increases from $\Pi = 0.01$ to 0.02 yet monotonically decreases from 0.02 to 0.1.
At saturation, $0 > \avgvx > v_{x,0} \approx -0.82\Pi\cs$ as more of the dust shifts towards the positive side of the initial equilibrium, while $\avgvz \approx v_{z,0} = 0$.
As with the gas (cf. Fig.~\ref{fig:dPduxz}), $\dPdvx$ becomes more Gaussian as $\Pi$ increases, while $\dPdvz$ appears somewhat uniform and much wider, but the width has no obvious trend with $\Pi$.
Moreover, as $\dPdvx$ becomes more negatively skewed as $\Pi$ decreases, we find fewer particles drifting inward between dense dust filaments in Fig.~\ref{fig:BA_snapshots} and in the videos for Case~BA (see `Data Availability' section).

As done for Case~AB, we show for Case~BA in the bottom rows of Figs.~\ref{fig:dPdvx} and \ref{fig:dPdvz} the average particle density $\rhop(v_x)$ and $\rhop(v_z)$, respectively.
Depending on the BA~model in Fig.~\ref{fig:dPdvx}, we find $\max(\rhop)$ between $0 \lessapprox v_x \lessapprox 0.3\Pi\cs$.
As $\Pi$ decreases, the $v_x$ with $\max(\rhop)$ for a given model more closely aligns with the peak of the corresponding $\dPdvx$ in the panel above.
Meanwhile, $\rhop$ decreases by two or more orders of magnitude as $v_x$ increases or decreases away from $\max(\rhop(v_x))$ \citep[cf.][fig.~16, Run~BA]{JohansenYoudin2007}.
For each BA~model in Fig.~\ref{fig:dPdvz}, we find the average density gradually increases with the magnitude of $v_z$ in either direction, leading to two maxima near the maximum speed reached by each model.
These maxima closely align with the variations seen in $\dPdvz$ above, but the alignment is more subtle with the velocity distribution $\dPduz$ for the gas (Fig.~\ref{fig:dPduxz}).
This can be seen in the coherent upward or downward motions of dense dust filaments seen dominating the saturation state in the videos (see `Data Availability' section) and as discussed in Section~\ref{sec:BA_morphology}.
Nevertheless, we note that for both Cases~AB and BA, the dust density $\rhop$ of cells with similar dust velocities varies widely, up to more than two orders of magnitude.

\begin{figure}
	\includegraphics[width=\columnwidth]{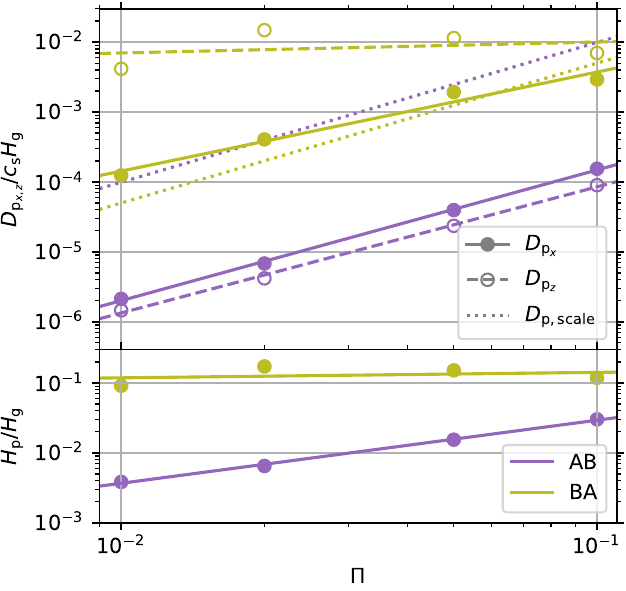}
    \caption{Diffusion coefficients (top) and estimated particle scale heights (bottom) of the dust at saturation as a function of $\Pi$ for Cases~AB ($\taus = 0.1$ with $\epsilon = 1.0$) and BA ($\taus = 1.0$ with $\epsilon = 0.2$).
    In both panels, different colours represent different cases, and solid or dashed lines represent the best-fitting power-law defined by equation~\eqref{eq:power_law}.
    In the top panel, solid lines and filled circles correspond to the radial component, dashed lines and open circles correspond to the vertical component, and dotted lines correspond to the scaling defined by equation~\eqref{eq:Dscale}.
    In the bottom panel, we plot the estimated particle scale height defined by equation~\eqref{eq:Hp}.}
    \label{fig:dust_motions}
\end{figure}

Next, we follow \citet[\S~4.1]{YangMacLowMenou2009} to compute the radial and vertical components of the diffusion coefficient $\Dpxz$ of the particles in each model.
The top panel of Fig.~\ref{fig:dust_motions} shows our results for $\Dpxz$ at saturation as a function of $\Pi$.
We note our $\Dpxz$ for the AB and BA~models when $\Pi = 0.05$ are in good agreement with those reported by \citet[table~3]{JohansenYoudin2007}.
In general, $\Dpxz$ increases as $\Pi$ increases, except in the vertical direction for Case~BA, where its dependence on $\Pi$ is weak with $\Dpz \sim 10^{-2}\cs\Hg$.
For all AB~models, $\Dpx \gtrsim \Dpz$, while $\Dpx < \Dpz$ for the $0.01 \leq \Pi \leq 0.1$ we investigated in Case~BA.

As was done for the Mach numbers and $\alpha$ parameters of the gas (Section~\ref{sec:gas_turbulence}), we fit the power law defined by equation~\eqref{eq:power_law} to the dust diffusion coefficients.
The best-fitting lines and parameters are shown in Fig.~\ref{fig:dust_motions} and listed in Table~\ref{tab:dust_fitting_parameters}, respectively.
Since the lengths and the velocities in the linear system for the streaming instability scale with $\Pi\Hg$ and $\Pi\cs$, respectively \citep{YoudinGoodman2005, YoudinJohansen2007}, the diffusion coefficient should scale with \citep[cf.][eqs.~4 and 5]{YoudinLithwick2007}
\begin{equation}
    \Dpscale \sim \frac{\Pi^2\cs\Hg}{1 + \taus^2}.
	\label{eq:Dscale}
\end{equation}
To compare with our best-fitting lines, we plot $\Dpscale$ for Cases~AB ($\taus = 0.1$) and BA ($\taus = 1.0$) in the top panel of Fig.~\ref{fig:dust_motions}.
For $\Dpxz(\Pi)$ in Case~AB, the power-law index $k \approx 2$, indicating the dust diffusion roughly scales with $\Dpscale \propto \Pi^2$.
Furthermore, the fact $\Dpxz \sim 10^{-2}\Dpscale$ agrees with the product of the characteristic turbulent eddy length scale $\sim 10^{-1}\Pi\Hg$, as estimated by the half width at half maximum of $\Rg$ in Fig.~\ref{fig:AB_avgRs_rad-prof}, and the characteristic turbulent velocity scale $\sigma_{v_{x,z}} \sim 10^{-1}\Pi\cs$ from Table~\ref{tab:velocities}.

\begin{table}
	\centering
	\caption{Parameters for the dust of the best-fitting power laws$\anote$ shown in Fig.~\ref{fig:dust_motions}.
	The columns correspond to (1) case name, (2)--(3) parameters for the radial component of the particle diffusion coefficient, (4)--(5) for the vertical component of the particle diffusion coefficient, and (6)--(7) for the estimated particle scale height$\bnote$.
        Diffusion coefficients and particle scale heights are in units of $\cs\Hg$ and $\Hg$, respectively.}
	\label{tab:dust_fitting_parameters}
    \begin{tabular}{cllllrl}
    \hline
    Case & 
    \multicolumn{2}{c}{$\Dpx(\Pi)$} &       
    \multicolumn{2}{c}{$\Dpz(\Pi)$} &       
    \multicolumn{2}{c}{$\Hp(\Pi)$} \\      

    &
    \multicolumn{2}{c}{$(\cs\Hg)$} &       
    \multicolumn{2}{c}{$(\cs\Hg)$} &       
    \multicolumn{2}{c}{$(\Hg)$}    \\      
    
    &
    \multicolumn{1}{c}{$a/10^{-2}$} & 
    \multicolumn{1}{c}{$k$} &              
    \multicolumn{1}{c}{$a/10^{-2}$} & 
    \multicolumn{1}{c}{$k$} &              
    \multicolumn{1}{c}{$a$} &              
    \multicolumn{1}{c}{$k$} \\             
         
    \multicolumn{1}{c}{(1)}  &
    \multicolumn{1}{c}{(2)}  &
    \multicolumn{1}{c}{(3)}  &
    \multicolumn{1}{c}{(4)}  &
    \multicolumn{1}{c}{(5)}  &
    \multicolumn{1}{c}{(6)}  &
    \multicolumn{1}{c}{(7)}  \\
    
    \hline       
    \multirow{1}{*}{AB} & 1.1 & 1.9 & 0.5 & 1.8 & 0.23 & 0.9 \\
    
    \multirow{1}{*}{BA} & 9.8 & 1.4 & 1.4 & 0.2 & 0.17 & 0.1 \\
    \hline
    \multicolumn{3}{l}{$\anote$ Defined by equation~\eqref{eq:power_law}} \\
    \multicolumn{3}{l}{$\bnote$ Defined by equation~\eqref{eq:Hp}} \\
    \end{tabular}
\end{table}

Akin to the results from our other analyses, the dust diffusion in Case~BA behaves rather differently than in AB or from what might be expected by the linear system of the streaming instability.
$\Dpx(\Pi)$ does not quite scale with $\Dpscale$, as the power-law index $k \approx 3/2$.
As particles are allowed to move freely in the vertical direction, $\Dpz$ appears insensitive to $\Pi$ with $k \approx 0$.
In addition, $\Dpxz > \Dpscale$ for $\Pi \lesssim 0.1$.
\cite{JohansenYoudin2007} attribute the strong vertical diffusion they also found in their Run~BA to the bulk vertical motion of dense dust filaments, which we describe in Sections~\ref{sec:BA_morphology}.
Given the particle stopping time and the velocity dispersion (Table~\ref{tab:velocities}), the implied characteristic diffusion length scale exceeds $O(\Pi\Hg)$ for BA~models.

By equating the vertical diffusion time-scale $\Hp^2/\Dpz$ with the sedimentation time-scale $(\taus + \taus^{-1})/\OmegaK$ \citep[][eq.~2]{YoudinLithwick2007}, we can estimate the particle scale height as \citep[][eq.~9]{YangZhu2021},
\begin{equation}
    \Hp \approx \sqrt{\frac{\Dpz}{\OmegaK}\left( \taus + \frac{1}{\taus} \right)} = \Hg\sqrt{\frac{\Dpz}{\cs\Hg}\left( \taus + \frac{1}{\taus} \right)}.
    \label{eq:Hp}
\end{equation}
The bottom panel of Fig.~\ref{fig:dust_motions} shows our results for $\Hp$ at saturation as a function of $\Pi$.
In addition, we fit the power law defined by equation~\eqref{eq:power_law} to our particle scale height estimates and show the best-fitting lines and parameters in Fig.~\ref{fig:dust_motions} and Table~\ref{tab:dust_fitting_parameters}, respectively.
As $\Dpz$ and $\Hp$ are correlated, $\Hp$ increases as $\Pi$ increases in Case~AB but remains almost constant at around $0.1\Hg$ in Case~BA.
We note that our $\Hp$ estimates for the AB and BA~models when $\Pi = 0.05$ are in fair agreement with those for similar stopping times in Models~Af and B by \citet[cf. fig.~12]{YangZhu2021}, despite the fact that their simulations contain multiple dust species.

In the vertically stratified models when $\Pi = 0.05$ by \cite{LiYoudin2021}, they directly measured the particle scale height of mono-disperse dust and reported its time average taken over the saturated `pre-clumping' phase (defined in their Section~3.2).
Referring to Table~\ref{tab:params} and Fig.~\ref{fig:dust_motions}, the effective metallicity $\Zeff = \epsilon\Hp / \Hg$ for our AB and BA~models when $\Pi = 0.05$ are $\Zeff \approx 0.015$ and $\approx 0.030$, respectively.
Compared to our AB~model, their models with $\taus = 0.1$ span $0.005 \leq Z \leq 0.01$.
Although these $Z < \Zeff$, the relative differences between $\Hp \approx 0.015\Hg$ estimated in our AB~model and the time-averaged $\Hp$ they measured in their models are at most 3\%, after converting from units of $\eta r$ in their Table~2 to $\Hg$ via equation~\eqref{eq:Pi}.
Compared to our BA~model, however, our estimated $\Hp \approx 0.15\Hg$ greatly exceeds the $0.006 \lessapprox \Hp \lessapprox 0.007$ measured in their $\taus = 1.0$ models.
Even though $\Hp$ increases as $Z$ increases from $0.004 \leq Z \leq 0.0075$, the enhanced turbulence at the mid-plane can make identifying a pre-clumping phase more difficult.
Indeed, their highest $Z = 0.01 \approx \Zeff/3$ model with $\taus = 1.0$ lacked a pre-clumping phase as particles concentrated too quickly, and they did not report $\Hp$.
Thus, it remains unclear if a fair comparison of $\Hp$ with our BA~model can be made.

\section{Discussion}
\label{sec:discussion}

\subsection{Implications for planetesimal formation and radial transport}
\label{sec:implications}

The measurement of dust diffusion in our models may help us understand why greater solid abundances may be needed to trigger strong particle clumping \citep[][fig.~2]{BaiStone2010L} and may lead to less efficient planetesimal formation \citep[][fig.~3]{AbodSimonLi2019} for stronger radial pressure gradients.
For both cases we have studied, the strength of the gradient directly affects dust diffusion driven by the streaming instability at non-linear saturation.
As detailed in Sections~\ref{sec:saturation_state} and \ref{sec:kinematics}, the dust velocity dispersions along each directional component increase as the gradient increases (Fig.~\ref{fig:dispersions} and Table~\ref{tab:time_averages}).
Furthermore, the dust diffusion coefficient increases with the gradient except in the vertical direction for the marginally-coupled case (Table~\ref{tab:dust_fitting_parameters} and the top panel of Fig.~\ref{fig:dust_motions}).
As discussed in \citet[][\S~3.3]{YangJohansen2014} and \cite{YangJohansenCarrera2017}, the strong radial concentration of dust into axisymmetric filaments by the streaming instability (with vertical sedimentation) resembles the process of traffic jams which collect upstream particles.
Stronger dust diffusion, however, should lower the concentration of these filaments, and hence a higher solid abundance to elevate the dust density should be needed to trigger gravitational collapse.
Moreover, even after gravitational collapse takes over, the characteristic size of the planetesimals becomes larger with stronger turbulent diffusion \citep{KlahrSchreiber2021}.
Hence, as the radial pressure gradient increases, it becomes difficult for planetesimals of larger and larger sizes to form, leading to less efficient planetesimal formation.

Similar to previous studies, we find turbulence driven by the streaming instability affects the radial drift of dust throughout the protoplanetary disc.
In particular, we find the instability should transport tightly-coupled particles more efficiently than without turbulence.
Their average velocity at saturation is double the drift rate at initial equilibrium and increases in linear proportion with the gradient strength (Table~\ref{tab:velocities}).
Thus, radial transport of these particles should occur in half the expected drift time-scale wherever the instability may develop, independent of the local magnitude of the (non-zero) gradient.

For marginally-coupled particles, on the other hand, we find somewhat different implications for global transport.
While the magnitude of their average radial velocity at saturation is less than that at initial equilibrium, it increases in super-linear proportion with the gradient strength.
As a result, their drift may gradually slow down as the gradient decreases on approach of a nearby pressure maxima, until the instability stops when the gradient reaches zero and they perhaps remain trapped (Section~\ref{sec:introduction}).
This reduction in their average speed may enhance the collecting capabilities of the characteristic radial traffic jams seen in this case (Section~\ref{sec:BA_morphology}), thereby increasing the local density of marginally-coupled particles (Fig.~\ref{fig:densities}), which in turn may lead to stronger clumping and more efficient planetesimal formation near pressure maxima.
Even though results from \cite{CarreraSimon2022} seem to have challenged this latter implication for pressure maxima, it remains unclear if their simulations satisfied the modelling requirements suggested by \citet[][\S\S~3.5.1, 4.1.2, and 4.1.4]{LiYoudin2021} to facilitate strong clumping.

Turbulent diffusion should also contribute to dust transport, but it is less important than radial drift on length-scales of interest.
We find that for a given gas scale height $\Hg$, the drift time-scale ($\Hg / |\avgvx|$) is at least two orders of magnitude faster than the diffusion time-scale ($\Hg^2 / \Dpx$; Fig.~\ref{fig:dust_motions}), as was found by \citet[][\S~5.2]{BaiStone2010}, \citet[][\S~6]{SchafferYangJohansen2018}, and \citet[][\S~4.6]{YangZhu2021}.
Note though, turbulent diffusion should remain important on much smaller scales.

\subsection{Comparisons with observations}
\label{sec:observations}

Kinematic analysis of the non-thermal broadening of molecular-line emissions can constrain the magnitude of turbulent gas motions in observed protoplanetary discs.
\cite{FlahertyHughesRosenfeld2015, FlahertyHughesRose2017} used several CO emission lines with various optical depths, probing layers of different heights, to find a characteristic Mach number (Section~\ref{sec:gas_turbulence}) $\Ma < 0.05$ in the outer disc of HD~163296.
As shown in Fig.~\ref{fig:gas_turbulence}, our measurements agree with this upper limit, except for vertical gas motions $\Ma_z$ in Case~BA when $\Pi = 0.1$.
Using the same methodology, \cite{FlahertyHughesTeague2018} found $\Ma < 0.08$ for TW~Hya, and \cite{FlahertyHughesSimon2020} found $\Ma < 0.08$ for MWC~480 and $\Ma < 0.12$ for V4046~Sgr.
By analysing CS, which is heavier and thus less sensitive to thermal broadening than CO, \cite{TeagueHenningGuilloteau2018} constrained $\Ma \lesssim 0.1$ across the entire disc of TW~Hya.
All of our measurements are in agreement with all of these subsequent findings.
For DM~Tau, on the other hand, \cite{FlahertyHughesSimon2020} found $0.25 < \Ma < 0.33$, which exceeds our direct measurements yet implies $\Pi \gtrsim 1$ from our our best-fitting power-laws (Table~\ref{tab:gas_fitting_parameters}, Columns~2 and 4) if the gas turbulence is driven by the streaming instability alone.
Such steep pressure gradients could trigger the Rossby wave instability \citep{LiFinnLovelace2000, OnoMutoTakeuchi2016, ChangYoudinKrapp2023}, which should in turn drive non-axisymmetric structures \citep{LyraLin2013, vanderMarelvanDishoeckBruderer2013}.
Therefore, other sources of turbulence are required to explain the higher Mach numbers observed \citep[][\S~5.2]{FlahertyHughesSimon2020}.
Lastly, \citet[][\S~4.5]{FlahertyHughesTeague2018} and \citet[][\S~3.2]{PinteTeagueFlaherty2022} note the temperature profiles assumed and the amplitude calibration of optically thick emission, respectively, represent significant sources of uncertainty in distinguishing thermal and non-thermal broadening.

By comparing millimetre-continuum and molecular-line emissions with radiative-transfer models, several studies have attempted to constrain the dust and gas scale heights of observed discs.
For their models to account for the well-observed gaps and bright rings in HL~Tau, \citet[][\S~6]{PinteDentMenard2016} found an upper limit of $\Hp \lessapprox 0.2\Hg$ for millimetre-sized grains at $r = 100$~au.
Integrating their dust-size distribution up to 3~mm, they also estimated a mid-plane dust-to-gas mass ratio of $\epsilon \gtrapprox 0.2$ at 100~au.
As shown in the bottom panel of Fig.~\ref{fig:dust_motions}, our dust scale-height estimates for both of our cases are consistent with their upper limit.
For HD~163296, \citet[][fig.~3]{IsellaGuidiTesti2016} inferred the scale heights of the CO gas and dust for $r \leq 300$~au in the disc, which our $\Hp \approx 0.1\Hg$ estimate for Case~BA agrees with well across almost the entire range.
For the outer ring at 100~au in this disc, \cite{DoiKataoka2021} found $\Hp < 0.11\Hg$, consistent with particle scale-height estimates for all of our models.
More recently, \cite{VillenaveStapelfeldtDuchene2022} found an upper limit for the dust scale height at $r \sim 100$~au in the highly inclined disc Oph~163131, where their new observations at high angular resolution revealed a clear outer dust ring.
Using the lowest limit from \cite{WolffDucheneStapelfeldt2021} for the gas scale height, this yields $\Hp \leq 0.08\Hg$, which is consistent with our results for Case~AB but in rough agreement with those for Case~BA.

The scale heights discussed above can be considered relatively thin, indicating some degree of settling to the mid-plane, but studies have also found larger values of $\Hp / \Hg$.
For HD~163296, \cite{OhashiKataoka2019} used the polarized emission to constrain grain sizes in their models and found $\Hp \lessapprox \Hg/3$ and $\approx 2\Hg/3$ at the respective gaps interior (48~au) and exterior (86~au) to the thick inner ring at 68~au, which \cite{DoiKataoka2021} estimated to have $\Hp > 0.84\Hg$.
Although the largest of these estimates exceeds ours (Fig.~\ref{fig:dust_motions}, bottom panel) by almost one order of magnitude (Case~BA) or more (Case~AB), our best-fitting power-law (Table~\ref{tab:dust_fitting_parameters}, Column~6) implies $\Pi \sim 1$ at the inner gap at 48~au.
If pressure maxima, where $\Pi \approx 0$, are indeed trapping dust to form the bright rings observed in HD~163296, e.g. at 68~au, stronger pressure gradients might be expected at the gaps in between.
For the edge-on disc IRAS04302+2247, \cite{VillenavePodioDuchene2023} found $0.1 < \Hp / \Hg < 0.9$ for millimetre-sized grains at 100~au, again exceeding the direct estimates from our models.
As argued by \cite{DullemondBirnstielHuang2018}, diffusive processes must contribute to the extended nature of these dust features as they act against (1) vertical settling into a thin layer at the mid-plane due to gravity and (2) radial trapping into a thin annulus at the local pressure maximum due to aerodynamics.
If the streaming instability alone is responsible for these observed thick features, our results suggest $\Pi \gtrsim 1$, hence other sources of vertical dust diffusion, such as the vertical shear instability \citep{SchaferJohansen2022, Lin2019, FlockTurnerNelson2020} and the magneto-rotational instability \citep{RiolsLesur2018, YangMacLowJohansen2018, HuLiZhu2022, XuBai2022}, should be considered.

To constrain the degree of dust--gas coupling amidst turbulent mixing, recent observational studies have reported the ratio $\alpha / \St$, where $\St$ is referred to as the Stokes number of the solid particle (Section~\ref{sec:dust}).
However, estimates of this quantity have assumed the dust diffusion driven by the gas turbulence are both homogeneous and isotropic, which may not be the case in theoretical \citep{JohansenKlahrMee2006} or observed protoplanetary discs \citep{DoiKataoka2021, WeberCasassusPerez2022, VillenaveStapelfeldtDuchene2022}.
Moreover, stand-alone $\alpha$ estimates of observed discs, when reported via comparisons with dust vertical settling or radial trapping models, must assume values for $\taus$ \citep[][\S~3.2]{PinteTeagueFlaherty2022}, which can differ between observational diagnostics \citep{UedaKataokaZhang2021}, and the gas density $\rhog$, which may be unknown.
To order unity agreement with a more precise formulation, the ratio of the gas diffusion coefficient to that for the dust is $D_\mathrm{g} / D_\mathrm{p} \sim 1 + \taus^2$ \citep[][eq.~5]{YoudinLithwick2007}.
Since we separately measure the radial and vertical components of the particle diffusion coefficient $\Dpxz$, we can compute
\begin{equation}
    \frac{\alpha_{x,z}}{\taus} = \frac{D_{\mathrm{g}_{x,z}}}{\cs\Hg\taus} \sim \frac{\Dpxz}{\cs\Hg}\left(\taus + \frac{1}{\taus}\right)
    \label{eq:alpha_xz}
\end{equation}
to characterise the effect of gas turbulence on dust diffusion.
With equation~\eqref{eq:alpha_xz}, $\alpha_{x,z} / \taus$ can be estimated from the top panel of Fig.~\ref{fig:dust_motions} by shifting values for Case~AB or BA upwards by one order of magnitude or by a factor of two, respectively.
Therefore, across the range of $10^{-2} \leq \Pi \leq 10^{-1}$ we study, $10^{-5} \lessapprox \alpha_{x,z} / \taus \lessapprox 10^{-3}$ for Case~AB, and $2 \times 10^{-4} \lessapprox \alpha_x / \taus \lessapprox 6 \times 10^{-3}$ and $\alpha_z / \taus \approx 2 \times 10^{-2}$ for Case~BA.

With the assumption of $\St = \taus$,\footnote{
As discussed in \citet[][\S~2]{YoudinLithwick2007}, the Stokes number $\St \equiv \tstop / \teddy = \taus / \taue$ relates the particle coupling $\taus$, defined by equation~\eqref{eq:taus}, to turbulent fluctuations, where the dimensionless eddy time $\taue \equiv \OmegaK \teddy$ characterises the effect of Keplerian shear on eddies with a turnover time of $\teddy$.
The contemporary conventional use of $\St = \taus$ implicitly assumes $\taue = 1$, but $\taue \neq 1$ can occur.}
we can compare $\alpha / \St$ estimates from observational studies that model the annular width of dust rings, resulting from radial diffusion, against $\alpha_x / \taus$.
\citet[][table~3]{DullemondBirnstielHuang2018}, for their lowest estimate for the ring at 77~au in Elias 24, and \citet[][table~1]{RosottiTeagueDullemond2020}, for the outer-most rings at 100 and 155~au in HD~163296, both found $\alpha_x / \taus \sim 10^{-2}$.
This estimate is about one order of magnitude larger than ours for our highest $\Pi = 0.1$ in Cases~AB and BA yet is consistent with our best-fitting power laws for $0.1 < \Pi < 1$ (Table~\ref{tab:dust_fitting_parameters}, Column~2).
Meanwhile, these studies also found $\alpha_x / \taus \sim 10^{-1}$ as their lowest estimates for rings in GW~Lup and HD~143006 \citep{DullemondBirnstielHuang2018} and as their general estimates for the inner-most ring at 68~au in HD~163296 and the two rings at 74 and 120~au in AS~209 \citep{RosottiTeagueDullemond2020}.
Judging by our best-fitting $\Dpx(\Pi)$ power laws for $\taus = 0.1$ and 1, this implies $\Pi \sim 1$, which again could trigger additional instabilities and contribute to higher turbulent $\alpha$ values for the same $\taus$.
Moreover, given the inherent degeneracy of $\alpha_{x,z} / \taus$, these higher values could also be accounted for by particles more tightly-coupled to the gas than those studied here, i.e. $\taus \ll 0.1$.

We can also compare $\alpha_z / \taus$ against studies that alternatively model the vertical scale height of dust rings that result from vertical diffusion.
\citet[][\S~5.2]{VillenaveStapelfeldtDuchene2022} found $\alpha/\taus < 6 \times 10^{-3}$ at 100~au in Oph~163131, and \citet{DoiKataoka2021} found $\alpha_z / \taus < 1.1 \times 10^{-2}$ for the thin outer ring at 100~au in HD~163296.
These upper limits agree with all of our estimates for Case~AB in Fig.~\ref{fig:dust_motions}.
Meanwhile, \citet[][\S~6.2]{OhashiKataoka2019} estimate $\alpha_z / \taus \lessapprox 0.1$ in the gap interior to the inner ring at 68~au also in HD~163296, which is consistent with those for both of our Cases~AB and BA.
However, they also found $\alpha_z / \taus \approx 0.8$ in the gap exterior to the same thick inner ring, at which \citet{DoiKataoka2021} found $\alpha_z / \taus > 2.4$.
These exceed our highest estimates by more than one order of magnitude in either of our cases or imply $\Pi \gg 1$ based on our best-fitting power law (\ref{tab:dust_fitting_parameters}, Column~4).
Furthermore, this $\alpha_z / \taus \gg \alpha_x / \taus \sim 10^{-1}$ estimate by \citep{RosottiTeagueDullemond2020} for the same thick inner ring at 68~au in HD~163296 quoted above.
Again, we note \cite{DoiKataoka2021, WeberCasassusPerez2022, VillenaveStapelfeldtDuchene2022} all suggest gas turbulence and dust diffusion are likely anisotropic in these observed discs, i.e. stronger radially than vertically.
Since we find the converse situation in Case~BA and at most a factor of two anisotropic difference in Case~AB, the streaming instability alone may not suffice to explain these observational findings.

\section{Conclusions}
\label{sec:conclusions}

In this work, we investigate the non-linear saturation of the streaming instability with a single dust species to quantify the dependence of dust--gas dynamics on the background radial pressure gradient.
To bridge the gap between mid-plane turbulence and vertical sedimentation, we focus on numerical simulations of a vertically-unstratified disc as the instability develops.
We survey four values of the dimensionless pressure gradient $\Pi$, spanning one order of magnitude from $0.01$ to $0.1$, and study two distinct combinations of the dimensionless stopping time $\taus$ and the dust-to-gas mass ratio $\epsilon$ and (Table~\ref{tab:params}).

Consistent with previous studies, the saturation state of the streaming instability with tightly-coupled particles is turbulent, consisting of numerous dust--gas vortices (Fig.~\ref{fig:AB_snapshots}).
Kinematic analyses of the gas and the dust (Sections~\ref{sec:saturation_state} and \ref{sec:kinematics}; the top panels of Figs.~\ref{fig:gas_turbulence} and \ref{fig:dust_motions}), show slightly stronger motions in the radial direction than in the vertical.
In general, we find dust and gas velocities (left-hand columns of Figs.~\ref{fig:dPduxz}, \ref{fig:dPdvx}, and \ref{fig:dPdvz}; Tables~\ref{tab:time_averages} and \ref{tab:velocities}) and the characteristic sizes of vortices (Fig.~\ref{fig:AB_avgRs_rad-prof}) scale in linear proportion with the pressure gradient.
Since the structures are smaller for weaker gradients, higher grid resolutions are required to trigger and study the instability (Appendix~\ref{sec:resolution_study}).
On the other hand, the distribution of gas densities shows a super-linear relationship with $\Pi$ in its width (Fig.~\ref{fig:densities}, top-right panel; Table~\ref{tab:time_averages}, Column~3).
Moreover, the dust density distribution (Fig.~\ref{fig:densities}, top-left panel) and the maximum dust concentrations reached (Table~\ref{tab:time_averages}, Column~11) seem largely insensitive to the gradient strength, and the reasons behind this remain unclear.

Also consistent with previous studies, the saturation state of the streaming instability with marginally-coupled particles consists of upward and downward moving patterns of dust filaments (Fig.~\ref{fig:BA_snapshots}).
Hence, gas and dust kinematics (Sections~\ref{sec:saturation_state} and \ref{sec:kinematics}; the top panels of Figs.~\ref{fig:gas_turbulence} and \ref{fig:dust_motions}), generally show much stronger motions in the vertical direction than in the radial.
Likely for the same reason, the dependence on the radial pressure gradient is more complex than its counterpart for tightly-coupled particles.
The magnitude of the average radial velocities for the gas and dust increases super-linearly with the gradient (Table~\ref{tab:velocities}, Columns~4 and 8), the corresponding radial dispersions scale linearly (Columns~5 and 9), and the vertical velocity dispersions scale sub-linearly except perhaps for $\Pi = 0.01$ (Columns~6 and 10).
As the gradient decreases, we find a decreasing radial separation and fewer particles drifting between filaments, along with increasing vertical segmentation and steeper tilts (Figs.~\ref{fig:BA_snapshots} and \ref{fig:BA_avgRs}).
Furthermore, the gas density dispersion (Table~\ref{tab:time_averages}, Column~3) scales linearly with the gradient, except perhaps for $\Pi = 0.1$, while the maximum dust concentration reached decreases by more than one order of magnitude between $\Pi = 0.01$ and $\Pi = 0.1$ (Column~11; Fig.~\ref{fig:densities}, bottom-left panel).

At the saturation state of the streaming instability, some of the properties that depend on the radial pressure gradient may be observable (Section~\ref{sec:observations}).
For the gas, we find the turbulent Mach number and the magnitude of the Reynolds stress $|\alpha|$ increase in linear proportion with $\Pi$ (Fig.~\ref{fig:gas_turbulence}).
Moreover, as $\Pi$ increases, the vertical scale height of particles should increase for those tightly coupled to the gas but remain about the same for those marginally coupled (Fig.~\ref{fig:dust_motions}, bottom panel).
These findings may help reconstruct the properties of observed discs where streaming turbulence dominates.

Finally, our results have important consequences for planetesimal formation and radial transport (Section~\ref{sec:implications}).
Except in the vertical direction for marginally-coupled particles, the increased dust diffusion from stronger pressure gradients (Fig.~\ref{fig:dust_motions}, top panel; Table~\ref{tab:dust_fitting_parameters}) should lower the concentration of filaments, potentially explaining previous findings from vertically-stratified simulations of the streaming instability: (1) the need for higher solid abundances to trigger strong particle clumping \citep{BaiStone2010L}; and (2) a reduced planetesimal formation efficiency \citep{AbodSimonLi2019}.
Furthermore, at non-linear saturation, the instability can radially transport tightly-coupled particles twice as fast than in laminar disks, independent of the gradient strength, while the radial drift of marginally-coupled particles slows down as the gradient decreases, which may lead to stronger clumping and more efficient planetesimal formation near pressure maxima.

\section*{Acknowledgements}

We appreciate all of the detailed and useful comments made by our reviewer.
We would also like to thank Andrew N. Youdin, Leonardo Krapp, Debanjan Sengupta, and Wladimir Lyra for their helpful comments on this work.
SAB acknowledges support by the National Aeronautics and Space Administration (NASA) under Grant No. 80NSSC20M0043.
CCY is grateful for the support by NASA TCAN program (grant number 80NSSC21K0497) and by the Munich Institute for Astro-, Particle and BioPhysics (MIAPbP) which is funded by the Deutsche Forschungsgemeinschaft (DFG, German Research Foundation) under 
Germany's Excellence Strategy -- EXC-2094 -- 390783311.
CCY and ZZ acknowledge the support by NASA via the Emerging Worlds program (grant numbers 80NSSC20K0347 \& 80NSSC23K0653) and the Astrophysics Theory Program (grant number 80NSSC21K0141).
Resources supporting this work were provided by the NASA High-End Computing (HEC) Program through the NASA Advanced Supercomputing (NAS) Division at Ames Research Center.

\section*{Data Availability}

Videos of the dust density field evolution can be accessed at \url{https://doi.org/10.6084/m9.figshare.c.6718221}.
Although the videos can be previewed from within a web browser, downloading them from the provided link and playing them locally will ensure they are viewed at the highest possible resolution and frame rate.
The remaining data underlying this article can be shared upon reasonable request to the corresponding author.



\bibliographystyle{mnras}
\bibliography{refs} 




\appendix

\section{Resolution study}
\label{sec:resolution_study}

For both Cases~AB and BA (Table~\ref{tab:params}), we conduct resolution studies from $256 \times 256$ cells up to at least $2048 \times 2048$ (Section~\ref{sec:model_setup}), maintaining an average of $n_p = 4$ particles per cell (Section~\ref{sec:dust}).
We find good agreement already between resolutions in some diagnostics or satisfactory convergence in others when $\Pi = 0.05$ (Section~\ref{sec:typical_rad_pres_grad}), a typical dimensionless value for the radial pressure gradient used in disc models (Section~\ref{sec:gas}).
With a much weaker $\Pi = 0.01$ on the other hand, we find various properties of the streaming instability in Case~AB require much higher resolutions to reach convergence (Section~\ref{sec:weaker_rad_pres_grad}).

\subsection{Fiducial Radial Pressure Gradient}
\label{sec:typical_rad_pres_grad}

\begin{figure*}
	\includegraphics[width=\textwidth]{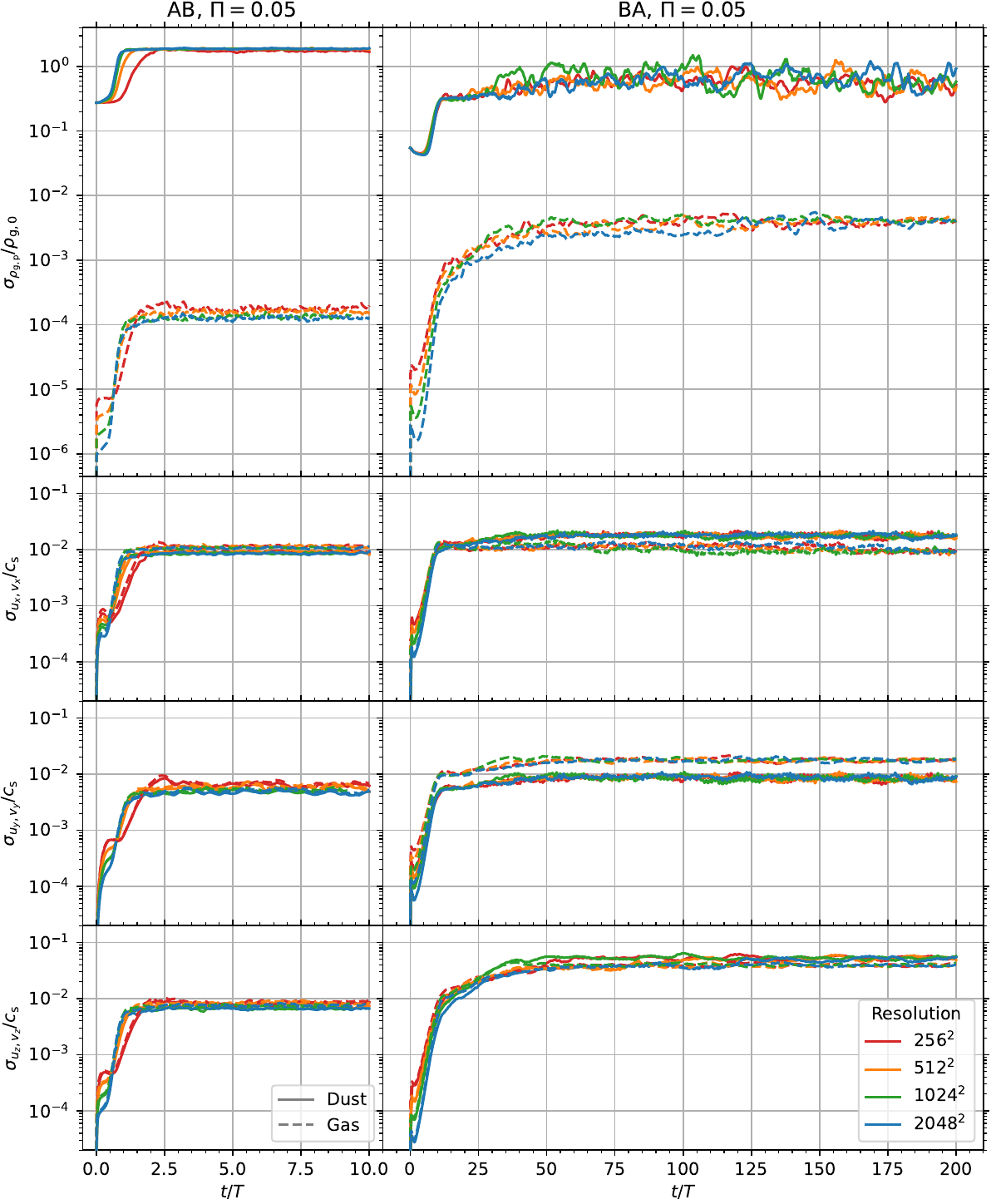}
    \caption{Similar to Fig.~\ref{fig:dispersions}, except with various resolutions for $\Pi = 0.05$.
    The different line colours represent models with different resolutions.}
    \label{fig:res_dispersions-0.05}
\end{figure*}

Several diagnostics for both Cases~AB and BA show good agreement between resolutions when $\Pi = 0.05$, the value most typically used in studies of the streaming instability (Section~\ref{sec:gas}).
Fig.~\ref{fig:res_dispersions-0.05} shows the evolution of density and velocity dispersions of the dust and gas (cf.\ Fig.~\ref{fig:dispersions}).
These dispersions all reach similar values at saturation across different resolutions, differing at most by a factor of two (e.g. $\sigma_{\rhog}$ in Case~AB and $\sigma_{\rhop}$ in Case~BA).
Models with higher resolutions show faster rates of increase prior to this state and reach saturation earlier (e.g. as seen in Case~AB for $t < 3T$) since faster-growing modes corresponding to higher radial and vertical wave numbers are resolved \citep[see][fig.~1]{YoudinJohansen2007}.

\begin{figure*}
	\includegraphics[width=\textwidth]{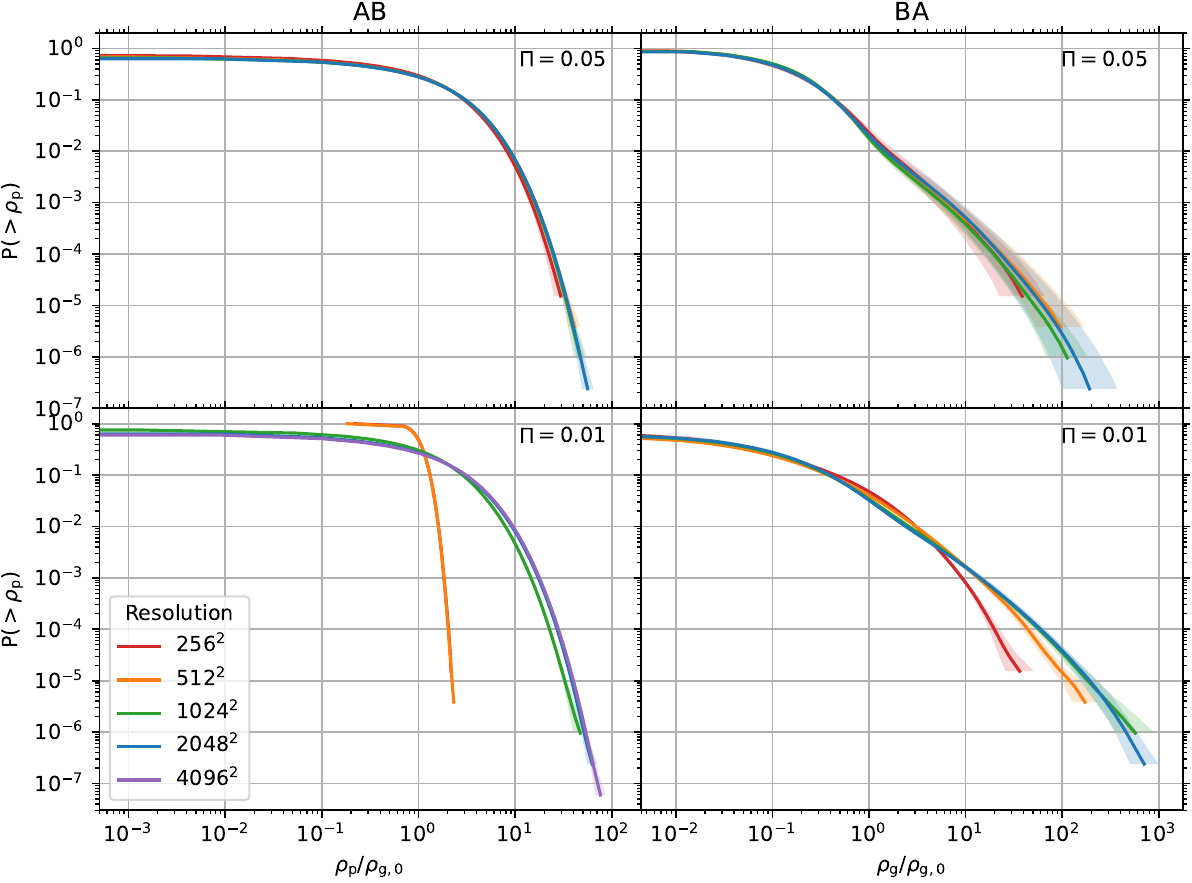}
    \caption{Similar to the left column of Fig.~\ref{fig:densities}, except with various resolutions for $\Pi = 0.05$ (top row) and 0.01 (bottom row).
   The different line colours represent models with different resolutions.}
    \label{fig:res_densities}
\end{figure*}

The time-averaged cumulative distribution functions for the dust density at various resolutions when $\Pi = 0.05$ also show good agreement between resolutions.
As shown in the top row of Fig.~\ref{fig:res_densities}, the distributions for each resolution overlap with one another to within their $1\sigma$ time variability.
If $\rhop$ is scaled by $\langle\rhop\rangle$ instead of $\rhogn$ via equation~\eqref{eq:epsilon}, these distributions are also in good agreement with those from the resolution studies for Runs~AB and BA conducted by \citet[][cf. fig.~6]{BaiStone2010S}.
We find less agreement with those by \citet[][cf. fig.~10]{Benitez-LlambayKrappPessah2019} and \citet[][cf. fig.~7]{HuangBai2022}, who instead used a multi-fluid approach and found a lack of convergence of the distribution function with resolution for run~AB.
Furthermore, since the distribution function shows consistency under the particle--mesh method in our models (Section~\ref{sec:numerical_method}), we find the maximum particle density $\max(\rhop)$ reached at saturation (i.e. the far-right, least-probable tails of the distributions in Fig.~\ref{fig:res_densities}) increases with increasing resolution, as previously reported by \citet[][supplemental \S\S~1.6.2 and 1.6.3]{JohansenOishiMacLow2007}, \citet[][\S~5.2]{BaiStone2010S}, and \citet[][\S~4.1]{JohansenYoudinLithwick2012}.
However, we also find these increases in $\max(\rhop)$ diminish with increasing resolution given the decreasingly small probability of finding high-density cells, indicating numerical convergence \citep[][cf. \S~3.2]{YangJohansen2014}.

\begin{figure*}
	\includegraphics[width=\textwidth]{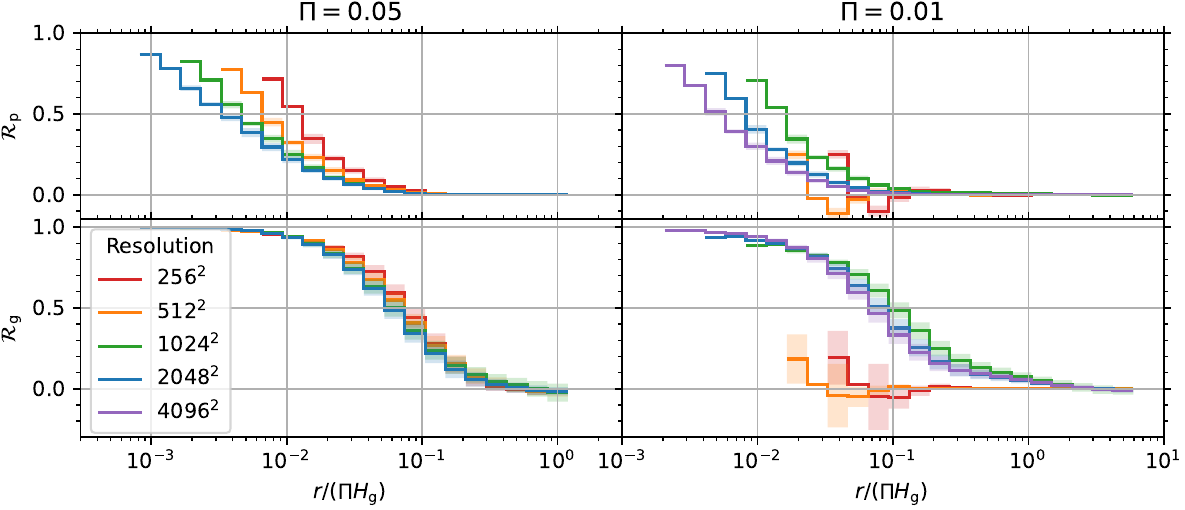}
    \caption{Similar to Fig.~\ref{fig:AB_avgRs_rad-prof}, except with various resolutions for $\Pi = 0.05$ (left column) and 0.01 (right column).
    The different line colours represent models with different resolutions.
    Radial bins are scaled by $\Pi$.}
    \label{fig:res_AB_avgRs_rad-prof}
\end{figure*}

As in Section~\ref{sec:AB_morphology}, we also compare time-averaged radial profiles of the normalised spatial auto-correlation (structure) functions, defined by equation~\eqref{eq:auto-correlation} of Section~\ref{sec:BA_morphology}, of the dust and gas density fields for Case~AB when $\Pi = 0.05$.
As shown in the bottom-left panel of Fig.~\ref{fig:res_AB_avgRs_rad-prof}, we find good agreement in the gas density profiles across resolutions and length scales, which overlap each other to within their $1\sigma$ time variability.
On the other hand, the dust density profiles (top-left panel) differ more towards smaller scales, but these differences diminish as the resolution increases, indicating numerical convergence.
Morphologically, the width of each profile roughly corresponds to the characteristic vortex size seen in snapshots of the dust density field (Section~\ref{sec:AB_morphology}; cf. Fig.~\ref{fig:AB_snapshots}).
Thus, the decrease in half width at half maximum of these profiles with an increase in resolution is consistent with the smaller sizes of dust vortices seen at higher resolutions in fig.~5 of \cite{BaiStone2010S} and fig.~8 of \cite{Benitez-LlambayKrappPessah2019}.
Numerical dissipation of the turbulent kinetic energy of particles at the smallest length scales \citep[][\S~4.2.1 and fig.~16]{SenguptaUmurhan2023} may contribute to this resolution effect on the dust morphology at saturation in Case~AB.

\subsection{Weaker Radial Pressure Gradient}
\label{sec:weaker_rad_pres_grad}

\begin{figure*}
	\includegraphics[width=\textwidth]{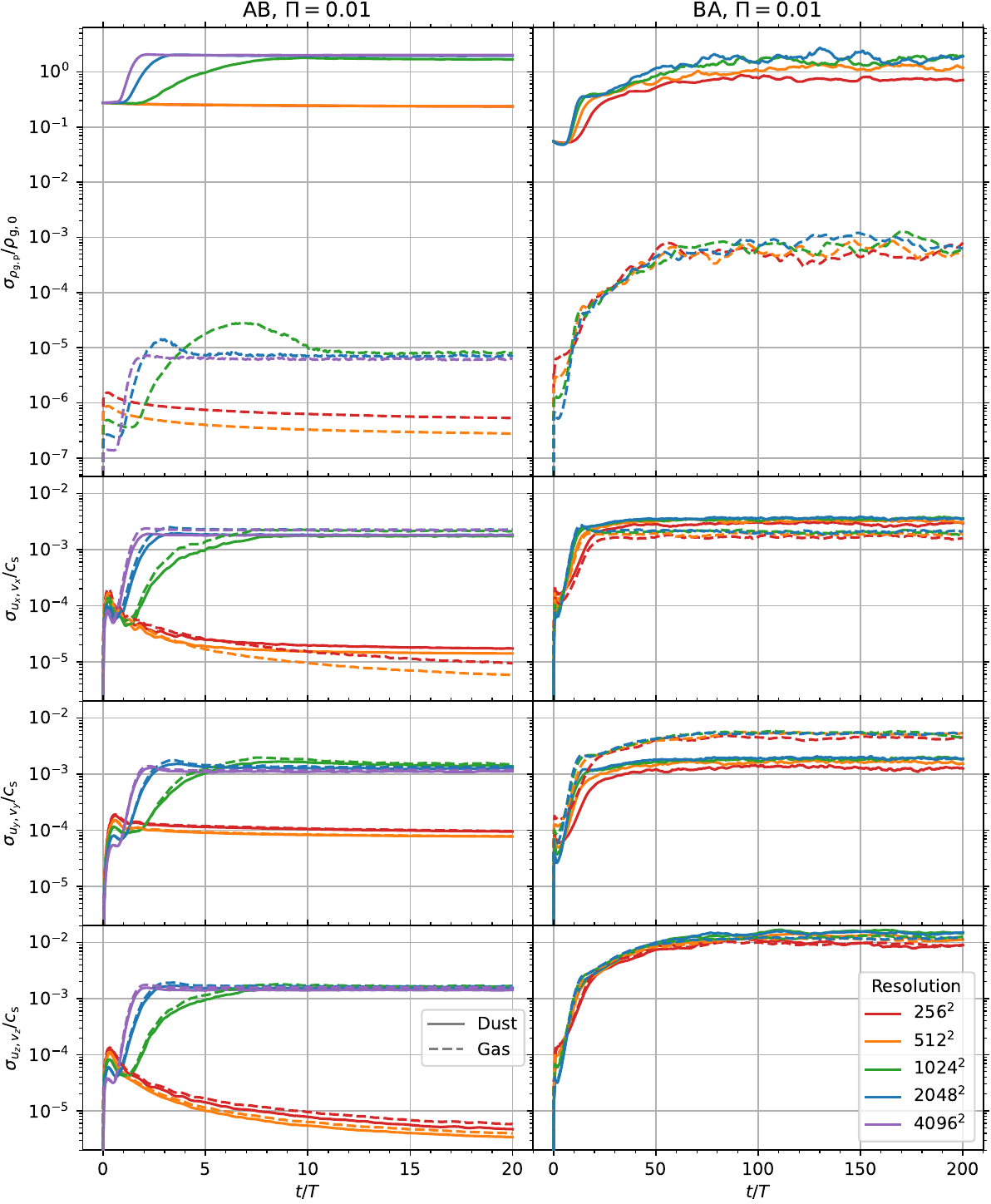}
    \caption{Similar to Fig.~\ref{fig:res_dispersions-0.05}, except for $\Pi = 0.01$.}
    \label{fig:res_dispersions-0.01}
\end{figure*}

In this section, we move to the much weaker pressure gradient of $\Pi = 0.01$.
For Case~BA, the diagnostics show either good agreement or satisfactory convergence between resolutions.
As shown in the right column of Fig.~\ref{fig:res_dispersions-0.01}, the density and velocity dispersions reach similar values at saturation between different resolutions, differing at most by a factor of a few (e.g. $\sigma_{\rhop}$).
Similar to when $\Pi = 0.05$ (Section~\ref{sec:typical_rad_pres_grad}), the dispersions increase faster and reach saturation sooner at higher resolutions (cf. Fig.~\ref{fig:res_dispersions-0.05}).
In addition, the time-averaged cumulative dust density distributions (Fig.~\ref{fig:res_densities}, bottom-right panel) begin to differ towards the lowest probabilities, but the differences diminish with increasing resolution (especially for the highest two resolutions we investigated), indicating convergence.

Case~AB, on the other hand, shows significant difficulty in obtaining a consistent saturation state.
The system never experiences linear growth nor reaches non-linear saturation at resolutions of $256^2$ or $512^2$.
As shown in the left column of Fig.~\ref{fig:res_dispersions-0.01}, their dust density dispersions maintain $\sigma_{\rhop} \approx 0.2\rhogn$, indicating a mostly uniform density field whose distributions (Fig.~\ref{fig:res_densities}, bottom-left panel) span at most one order of magnitude.
These two models likely cannot resolve the fast-growing modes for Case~AB ($\taus = 0.1$ with $\epsilon = 1.0$) when $\Pi \equiv \eta r/\Hg = 0.01$ (equation~\eqref{eq:Pi}), which can be located from the growth rate map in the bottom-centre panel of fig.~1 of \cite{YoudinJohansen2007}.
Moreover, radial profiles of the normalised spatial auto-correlation functions of their dust and gas density fields (Fig.~\ref{fig:res_AB_avgRs_rad-prof}, right column) show the presence of only weak perturbations, which can also be seen in the available videos (see `Data Availability' section).

At much higher resolutions, the diagnostics for Case~AB show either good agreement or satisfactory convergence.
Since we find the system only triggers linear growth at a resolution of $1024^2$ or higher, we run one additional model at $4096^2$.
The dispersions for these models at or above $1024^2$ (Fig.~\ref{fig:res_dispersions-0.01}, left column) increase faster and reach saturation sooner at higher resolutions.
Moreover, the model with a resolution of $1024^2$ reaches saturation only after $t = 10T$ (e.g. $\sigma_{\rhog}$).
Thus, for this resolution study alone, we extend $\tlim$ to $20T$ (cf. Table~\ref{tab:params} in Section~\ref{sec:model_setup}) and average the remaining diagnostics for the saturation state from $t = 15T$ to $20T$ (cf. Section~\ref{sec:saturation_state}).
The bottom-left panel of Fig.~\ref{fig:res_densities} shows satisfactory convergence of the dust density distribution with increasing resolution.
Similar to our findings when $\Pi = 0.05$ (Section~\ref{sec:typical_rad_pres_grad}), the gas density profiles for the three highest resolutions overlap each other to within their $1\sigma$ time variability (Fig.~\ref{fig:res_AB_avgRs_rad-prof}, bottom-right panel).
Also similar to when $\Pi = 0.05$, we find the differences between dust density profiles (top-right panel) towards the smallest scales diminish with increasing resolution for these three highest models, indicating convergence.
In conclusion, we recommend a minimum resolution of $2048 \times 2048$ for Case~AB when $\Pi = 0.01$.


\bsp	
\label{lastpage}
\end{document}